\documentclass{article}

\usepackage{latexsym}
\usepackage{amsmath}
\usepackage{amssymb}
\usepackage{amsthm}
\usepackage{amscd}

\usepackage{graphicx}
\usepackage{subfigure}
\usepackage{psfrag}

\usepackage{bm}
\usepackage{cancel}

\newcommand{\textfrac}[2]{{\textstyle \frac{#1}{#2}}}

\newtheorem{Theorem}{Theorem}
\newtheorem{Corollary}[Theorem]{Corollary}

\title{Dynamics of Bianchi type I elastic spacetimes}
\author{Simone Calogero\footnote{E-Mail: calogero@mct.uminho.pt}\\[0.2cm]
Departamento de Matem\'atica para a Ci\^encia e Tecnologia\\
Campus de Azur\'em da Universidade do Minho\\
4800-058 Guimar\~aes, Portugal\\[0.5cm]
J.~Mark Heinzle\thanks{E-Mail: Mark.Heinzle@univie.ac.at}\\[0.2cm]
Gravitational Physics\\ Faculty of Physics, University of Vienna\\
A-1090 Vienna, Austria}
\date { }

\begin{document}

\maketitle
\begin{abstract}
We study the global dynamical behavior of spatially homogeneous 
solutions of the Einstein equations in Bianchi type I symmetry,
where we use non-tilted elastic matter as 
an anisotropic matter model that naturally generalizes perfect fluids.
Based on our dynamical systems formulation of the equations we are
able to prove that
(i) toward the future all solutions isotropize; 
(ii) toward the initial singularity all solutions display oscillatory behavior;
solutions do not converge to Kasner solutions but oscillate between
different Kasner states. This behavior is associated with energy
condition violation as the singularity is approached.
\end{abstract}

\section{Introduction}
\label{intro}

Understanding the dynamics of cosmological models 
is one of the main goals of theoretical cosmology. 
It is generally conceded
that the analysis of generic cosmological models 
(which are space-times without symmetries) 
is exceedingly difficult; in fact,
at present a mathematically rigorous treatment seems to be 
out of reach.
Nevertheless, heuristic and numerical 
studies that have been performed over the years 
have resulted in the formulation of 
a number of conjectures on the asymptotic dynamics of 
generic cosmologies,
see~\cite{HUR} and references therein.
In particular, it is conjectured that the
generic singularity is intimately connected
with (in fact, actually built on) 
the dynamics of 
spatially homogeneous cosmologies.


Spatially homogeneous cosmological models have been analyzed 
intensively over the years, so that
both the asymptotics toward the initial singularity and the
asymptotics in the regime of infinite expansion (with infinitely diluted
matter) are well understood; see~\cite{WE} for a review.
Most of the results concern solutions of 
the Einstein equations coupled to a perfect fluid which is usually 
assumed to obey a linear equation of 
state.
However, it is unclear in general, how robust these 
results are under a change of the matter model.
For example, it was shown 
in~\cite{HU} that the structure of the initial singularity for 
collisionless matter is considerably different from that of a perfect fluid 
already for models of Bianchi type~I. 
It is assumed that such different behavior stems from 
the anisotropic character of the 
stress-energy tensor~\cite{R}.   


In this paper we investigate the global dynamics of 
spatially homogeneous solutions of Bianchi type~I 
with anisotropic \textit{elastic matter}.
Elastic matter is described by the general relativistic theory of elasticity 
that was formulated by Carter/Quintana in~\cite{CQ} and further elaborated by Kijowski/Magli~\cite{KM}, 
Beig/Schmidt~\cite{BS} and Karlovini/Samuelsson~\cite{KS};
see also~\cite{P, T,MWP}. For very recent work
on the static Einstein-elastic matter equations see~\cite{ABS}.
Our choice of matter model is 
motivated by the fact that elasticity theory offers a natural way of generalizing
perfect fluids to a class of anisotropic phenomenological matter models
without the need to resort to \textit{ad hoc} 
assumptions on the expression of the anisotropic pressures.
We consider elastic matter with a simple constitutive equation (Lagrangian)
that leads to a stress-energy tensor of the form
\begin{equation}
T_{\mu\nu}=T_{\mu\nu}^{\rm fluid}+b\,\Pi_{\mu\nu}\:,
\end{equation}
where $T_{\mu\nu}^{\rm fluid}$ is the stress-energy tensor of a perfect fluid 
with linear equation of state 
and $b$ is a constant which modulates the contribution of the anisotropic stress tensor
$\Pi_{\mu\nu}$. 

The main results of the paper are the following:
Toward the future we observe isotropization of models.
All solutions resemble
infinitely diluted perfect fluid solutions in the asymptotic regime,
which is in accord with physical intuition.
Toward the initial singularity the behavior of 
Bianchi type I models with elastic matter is significantly 
different from the behavior of perfect fluid models.
We prove that the behavior toward the singularity is \textit{oscillatory}.
In particular, there does not exist any solution
that converges to a Kasner solution (vacuum solution).
In the LRS case 
solutions oscillate between two different Kasner states
(the Taub solution and the non-flat LRS solution);
in the general case, the solution undergoes
a (probably infinite) sequence of Kasner states (vacuum states)
as the singularity is approached.
This behavior is intimately connected with energy condition
violation.

The paper is organized as follows. In Section~\ref{elasticspacetimes} 
we briefly outline the derivation of the stress-energy tensor 
for elastic matter and introduce 
the class of diagonal Bianchi type~I solutions of the Einstein equations which will be 
the subject of our analysis. In Section \ref{dynamicalsystem} we reformulate the 
Einstein equations with elastic matter for diagonal Bianchi type~I solutions as a 
reduced dimensionless dynamical system on a compact state space. 
Section~\ref{globaldyn} 
contains the basic qualitative results on the global dynamics of solutions. 
In Section~\ref{LRSsolutions}
we specialize to the LRS (locally rotationally symmetric) case;
the reduced number of degrees of freedom permits a comprehensive
analysis of the past attractor. 
In Section~\ref{antiLRSsolutions} we discuss another (non-generic) 
subclass of solutions (``anti-LRS solutions'') whose behavior
resembles that of LRS solutions.
Finally, the most difficult problem is
addressed in Section~\ref{past}:
We present a detailed analysis of the past attractor of 
the full system and the associated past asymptotic behavior of
generic solutions.




\section{Bianchi type~I elastic spacetimes}
\label{elasticspacetimes}

We begin this section with an introduction to the general relativistic 
theory of elasticity.
However, since the only object of the theory used in this paper
is the stress-energy tensor $T_{\mu\nu}$ for elastic matter,
we shall restrict ourselves to a discussion of the basic concepts 
needed for the derivation of $T_{\mu\nu}$.
Comprehensive presentations of relativistic elasticity can be 
found in the references listed at the end of the paper. 
Most of the conventions we adopt, in particular those 
for the definition of the shear scalar and the 
elastic equation of state, are taken from~\cite{KS}.
(An option for the reader who is not interested in the
derivation of $T_{\mu\nu}$, is to simply take $T_{\mu\nu}$ as 
a given anisotropic stress-energy tensor and to 
proceed to the discussion of the Einstein equations in Bianchi type~I,
see~\eqref{BianchiI} together with~\eqref{einstein} and~\eqref{stressE}.)

Let $(M,\bar{g})$ denote the space-time, i.e., 
a four-dimensional manifold $M$ with Lorentzian metric $\bar{g}$ that
is time-orientable; local coordinates on $M$ are given by $x^\mu$, $\mu = 0, \ldots,3$.
The \textit{material space} (or body manifold) 
$(N,\gamma)$ is a three-dimensional Riemannian manifold;
local coordinates are $X^A$, $A=1,2,3$.
Points in the material space identify particles of the material (in the continuum limit),
where 
$\gamma$ measures the distance between the particles 
in the natural (unstrained) state of the matter. In the applications, $\gamma$ is usually chosen to be flat.
The \textit{configuration function} 
$\psi$ is defined to be a (smooth) map
\[
\psi:M\to N\,,\qquad\quad x^\mu\mapsto X^A = \psi^A(x^\mu)\:,
\]
such that the kernel of the \textit{deformation gradient} 
$T\psi: TM \rightarrow TN$ is generated by a (future-directed unit) timelike 
vector field $u$, i.e., $\ker T\psi = \langle u \rangle$ or $u^\mu \partial_\mu \psi^A = 0$.
The vector field $u$ is the matter four-velocity; by construction, $\psi^{-1}(p)$ 
(i.e., the world-line of the particle $p \in N$) is an integral curve.

We define two metrics 
on the orthogonal complement $\langle u \rangle^\perp$ of $u$ in $TM$ (which coincides
with $T\Sigma$ if $u$ is hypersurface orthogonal and thus generates a family of spacelike
hypersurfaces $\Sigma\subset M$).
The Riemannian metric induced by $\bar{g}$ we denote by $g$:
\[
g_{\mu\nu}=\bar{g}_{\mu\nu}+u_\mu u_\nu\,.
\]
The pull-back of the material metric by the map $\psi$, i.e., $\psi^*(\gamma)$, 
is called the \textit{relativistic strain tensor} $h$:
\[
h_{\mu\nu}=\partial_\mu\psi^A\partial_\nu\psi^B\,\gamma_{AB}\:;
\]
since $h_{\mu\nu}u^\mu=0$, it is a metric in $\langle u \rangle^\perp$; since
$\mathcal{L}_u h_{\mu\nu}=0$, it is constant along the matter flow.
The metric $h_{\mu\nu}$ on  $\langle u \rangle^\perp$ is Riemannian, hence $h^\mu_{\ \nu}$ has three positive
eigenvalues $h_1$, $h_2$, $h_3$. 

The material is unstrained at $x$ iff $g_{\mu\nu}(x)=h_{\mu\nu}(x)$. The scalar quantity
\[
n=\sqrt{{\rm det}_gh}=\sqrt{h_1h_2h_3}
\]
is the \textit{particle density} of the material. 
This interpretation is justified by virtue of the continuity equation 
\[
\nabla_\mu\left(nu^\mu\right)=0\:.
\]
A specific choice of elastic material is made 
by postulating a \textit{constitutive equation}, i.e., the 
functional dependence of the (rest frame) energy density $\rho$ of 
the material on the configuration map, the deformation gradient 
and the space-time metric. An important class of materials is the 
one for which this functional dependence enters only through 
the principal invariants of the strain 
tensor. In this case we have
\begin{equation}\label{lagrangian}
\rho=\rho(q_1,q_2,q_3),
\end{equation}
where 
\[
q_1=\mathrm{tr}\, h,\qquad q_2=\mathrm{tr}\left(h^2\right),\qquad q_3=\mathrm{tr}\left(h^3\right)\:;
\] 
since $n^2=(q_1^3-3q_1q_2+2q_3)/6$,
one of the invariants $q_i$ can be replaced by the particle density $n$.
The materials described by~\eqref{lagrangian} 
generalize the class of isotropic, homogeneous, 
hyperelastic materials from the classical theory of elasticity, 
see~\cite{MH}. 
In many 
astrophysical applications (e.g., for the description of the solid 
crust of neutron stars), the effect of very large strains can be 
modeled by an elastic material in the \textit{quasi Hookean approximation}~\cite{CQ}. 
This corresponds to a constitutive equation $\rho$ that (i) depends on $n$; (ii) depends 
linearly 
on a quadratic invariant of the strain; (iii) has an absolute minimum at zero strain. 
Following~\cite{KS} 
we choose the quadratic strain invariant to be the \textit{shear scalar}, which is given by  
\begin{subequations}
\begin{align}
s^2 & =\frac{1}{36}\left[n^{-2}\left(q_1^3-q_3\right)-24\right]\,, 
\intertext{or, in terms of the eigenvalues $h_1$, $h_2$, $h_3$,} 
\label{shearscalar}
s^2 & =\frac{1}{12}\left[
\left(\sqrt{\frac{h_1}{h_2}}-\sqrt{\frac{h_2}{h_1}}\right)^2+
\left(\sqrt{\frac{h_1}{h_3}}-\sqrt{\frac{h_3}{h_1}}\right)^2+
\left(\sqrt{\frac{h_2}{h_3}}-\sqrt{\frac{h_3}{h_2}}\right)^2\right]\,.
\end{align}
\end{subequations}
Evidently, $s^2$ is non-negative, and $s^2=0$ (no shear) iff $h_{\mu\nu} \propto g_{\mu\nu}$ (or equivalently, $h_1 = h_2 = h_3$).

In this paper we shall consider a constitutive equation of the form
\begin{equation}\label{constitutiveeq}
\rho= \check{\rho}(n)+\check{\mu}(n) s^2\,,
\end{equation}
where $\check{\rho}(n)$ 
is the \textit{unsheared energy density} and $\check{\mu}(n)$ the \textit{modulus of rigidity}.
The stress-energy tensor associated with these 
materials is obtained as the variation with respect to the 
space-time metric of the matter action $S_M=-\int\sqrt{|\bar{g}|}\,\rho$. 
The result is given in~\cite[Sec.~6]{KS} and reads
\begin{subequations}\label{Tmunu}
\begin{align}
\label{stressenergytensor}
\bar{T}_{\mu\nu}=\rho\, u_{\mu} u_{\nu} \, +\:\, & T_{\mu\nu}\,, \\[1ex]
\label{anisotropicstress}
\text{where}\quad\:
T_{\mu\nu}\, = \:\, & p\,\,g_{\mu\nu}+
\frac{1}{6}\frac{\check{\mu}}{n^2}
\left[\frac{1}{3}\left({\rm tr}(h^3)-({\rm tr} h)^3\right)g_{\mu\nu}+({\rm tr}h)^2h_{\mu\nu}-(h^3)_{\mu\nu}\right].
\end{align}
\end{subequations}
Here $p$ is the isotropic (component of the) pressure, which is given by
\begin{equation}\label{isopressure}
p=\check{p}(n)+ \check{\nu}(n) s^2\,, \qquad \text{where}\quad 
\check{p}=n^2\frac{d}{dn}\left(\frac{\check{\rho}}{n}\right)\,,\quad
\check{\nu} = \left(n\frac{d\check{\mu}}{dn}-\check{\mu}\right)\,.
\end{equation}
The principal pressures $p_i$ (which are the [non-zero] eigenvalues of $T^\mu_{\ \nu}$)
are thus of the form $p_i = p + \delta p_i$; for an unstrained configuration, $p_i = p$, $i=1,2,3$.
For $\check{\mu}=0$ (or $s^2=0$), 
the elastic material reduces to a perfect fluid 
with stress-energy tensor $\bar{T}_{\mu\nu} = \rho u_\mu u_\nu + p g_{\mu\nu}$,
energy density $\rho = \check{\rho}$ and pressure $p =\check{p}$.

It remains to specify the functions $\check{\rho}$ and $\check{\mu}$ 
in the constitutive equation~\eqref{constitutiveeq}. 
We postulate 
a linear equation of state between the unsheared pressure $\check{p}$ and the unsheared energy density $\check{\rho}$,
\begin{alignat*}{3}
\check{p} & = a\check{\rho} & &  \qquad (a \in [-1,1]) \,,& & 
\intertext{and a linear equation 
of state between the modulus of rigidity $\check{\mu}$ and the unsheared pressure $\check{p}$,}
\check{\mu} & =b\,\check{p} & & \qquad (a b \geq 0)\:. & & 
\intertext{By~\eqref{isopressure} this is equivalent to setting}
\check{\rho} & = \rho_0 n^{a+1}\,, & & 
\qquad \check{\mu}= \rho_0 a b\, n^{a+1} & & \qquad (|a| \leq 1, \: a b \geq 0)
\end{alignat*}
for some constant $\rho_0>0$.
Accordingly,
\begin{equation}\label{rho}
\rho=\rho_0 n^{a+1}\left(1+ab\, s^2\right),\qquad\qquad p=a\rho\,.
\end{equation}
Since for an unstrained material $\rho = \check{\rho}$ and
$p_i = p = \check{p}$ hold, $i=1,2,3$, the bound $|a| \leq 1$ ensures that the 
dominant energy condition $|p_i| \leq \rho$
is satisfied for an unstrained configuration.
Furthermore, $a b \geq 0$ guarantees that the energy density is positive for 
all values of the shear scalar $s^2$ and 
has a minimum at zero shear.
When $b = 0$, the modulus of rigidity $\check{\mu}$ vanishes and the elastic matter
reduces to a perfect fluid with a linear equation of state $p = a \rho$;
the condition $|a| \leq 1$ ensures that the dominant energy condition $|p| \leq \rho$
is satisfied for this perfect fluid.
When $a=0$ (so that $p = 0$), the choice of $b$ is irrelevant, since $a b = 0$; 
this is clear because shear cannot occur for dust.
Henceforth, unless stated otherwise, 
by elastic matter we will always mean matter with constitutive equation~\eqref{rho},
where $a \in [-1,1]$ and $a b > 0$. 

Consider now a homogeneous space-time $(M,\bar{g})$ of Bianchi type I, i.e.,
\begin{equation}\label{BianchiI}
\bar{g}_{\mu\nu} dx^\mu dx^\nu =-dt^2+g_{ij}(t)dx^idx^j\:,
\end{equation}
where $g_{ij}(t)$, $i,j=1,2,3$, is a family of Riemannian metrics
that is induced on the spatially homogeneous hypersurfaces $t=\mathrm{const}$.
Let $\psi^A(t,x^i):(M,\bar{g}) \rightarrow (N,\delta_{AB})$ 
be a material configuration and 
$\partial_\mu\psi^A$ the corresponding deformation gradient. 
Compatibility with Bianchi type I symmetry forces the deformation gradient 
$\partial_\mu\psi^A$ and thus the matter four-velocity $u^\mu$ to be 
functions of $t$ only. 
We assume non-tilted matter: $u^\mu$ is orthogonal to the surfaces $t=\mathrm{constant}$,
i.e., $u^\mu = \partial_t$. 
This implies that $0=u^\mu \partial_\mu \psi^A = \partial_t\psi^A$
and thus $\partial_t \partial_\mu \psi^A = 0$; 
hence $\partial_\mu \psi^A$ is constant with $\partial_0 \psi^A= 0$. 
For the strain tensor we find
\[
h_{00}=h_{0k}=0,\qquad h_{ij}=\delta_{AB} \partial_i \psi^A \partial_j \psi^B = \mathrm{const}\:;
\] 
since $h^i_{\:j} = g^{i k} h_{jk}$, the matrix $h^i_{\:j}$ is time-dependent
as are its eigenvalues $h_1$, $h_2$, $h_3$.
It follows from~\eqref{Tmunu} that
\begin{equation}\label{T00ij}
\bar{T}_{00}=\rho,\qquad\bar{T}_{0k}=j_k = 0,\qquad \bar{T}_{ij}=T_{ij}\,
\end{equation}
where $T_{ij}$ is given in terms of $h_{ij}$ via~\eqref{anisotropicstress}.

The Einstein equations, in units $c = 1 = 8 \pi G$,
decompose into the momentum constraint $j_k =0$, 
which is automatically satisfied by~\eqref{T00ij}, 
the Hamiltonian constraint
\begin{subequations}\label{einstein}
\begin{equation}\label{constraint}
(\mathrm{tr}k)^2-k^i_{\ j}k^j_{\ i}-2\rho=0\:,
\end{equation}
and the evolution equations
\begin{equation}\label{evolution}
\partial_t g_{ij}=-2k_{ij}\quad\partial_{t}k^i_{\ j}=({\rm tr}k)k^i_{\ j}-T^i_{\ j}+\frac{1}{2}\delta^i_{\ j}(T^k_{\ k}-\rho).
\end{equation}
\end{subequations}
Here, $k_{ij}$ is the second fundamental form of the surfaces $t=\mathrm{constant}$; 
Latin indexes are raised and lowered with $g_{ij}$.

The Cauchy data associated with this initial value problem
is given by $g_{ij}(0)$, $k^i_{\ j}(0)$; in addition we prescribe $h_{ij} = h_{ij}(0) =\mathrm{const}$.
Without loss of generality we can assume that $g_{ij}(0)$ and $k^i_{\ j}(0)$
are diagonal 
(by choosing coordinates adapted to an orthogonal basis of eigenvectors of $k^i_{\ j}(0)$).
Furthermore we impose the condition that $h_{ij}$ is diagonal; 
in particular, by rescaling the spatial coordinates, we can assume $h_{ij}=\delta_{ij}$.
This assumption is consistent with the evolution equations:
Since the off-diagonal elements of the tensor $T^i_{\ j}$ form an homogeneous polynomial 
in $h^i_{\ j} = g^{i k} h_{j k}$, $i\neq j$, 
it follows from the evolution equations~(\ref{evolution}) 
that $(g_{ij}, k^i_{\ j}, h^i_{\ j})$ remain diagonal for all times. 
Henceforth, we will restrict our attention
to this special class of solutions of the equations~\eqref{einstein}, which are referred to
as \textit{diagonal models}. 

From $h^i_{\ j}= g^{ik} h_{k j} = \mathrm{diag}(g^{11},g^{22},g^{33})=\mathrm{diag}(h_1,h_2,h_3)$ 
we conclude that
\begin{equation}
s^2=\frac{1}{12}
\left[\frac{g^{11}}{g^{22}}+\frac{g^{22}}{g^{11}}+\frac{g^{11}}{g^{33}}+\frac{g^{33}}{g^{11}}+\frac{g^{22}}{g^{33}}+\frac{g^{33}}{g^{22}}-6\right],
\end{equation}
cf.~(\ref{shearscalar}), which can be inserted into~\eqref{rho}, i.e.,
\begin{subequations}\label{stressE}
\begin{equation}\label{rhoagain}
\rho = \rho_0\, (g^{11} g^{22} g^{33})^{(a+1)/2} \,( 1 + a b s^2)\,,\qquad (|a|\leq 1, ab >0)\,,
\end{equation}
to yield $\rho$ as a function of $g^{11}$, $g^{22}$, $g^{33}$.
Moreover, from~(\ref{anisotropicstress}) we find
\begin{align}
T^1_{\ 1}& =p+\frac{1}{6}\check{\mu}\left(\frac{g^{11}}{g^{33}}-\frac{g^{33}}{g^{11}}+\frac{g^{11}}{g^{22}}-\frac{g^{22}}{g^{11}}\right),\\
T^2_{\ 2}& =p+\frac{1}{6}\check{\mu}\left(\frac{g^{22}}{g^{11}}-\frac{g^{11}}{g^{22}}+\frac{g^{22}}{g^{33}}-\frac{g^{33}}{g^{22}}\right),\\
T^3_{\ 3}& =p+\frac{1}{6}\check{\mu}\left(\frac{g^{33}}{g^{22}}-\frac{g^{22}}{g^{33}}+\frac{g^{33}}{g^{11}}-\frac{g^{11}}{g^{33}}\right),
\end{align}
\end{subequations}
where $p = a \rho$ and $\check{\mu} = \rho_0 a b (g^{11}g^{22}g^{33})^{(a+1)/2}$ and are thus functions of $g^{11}$, $g^{22}$, $g^{33}$.
In the following we analyze the equations~\eqref{einstein} with anisotropic 
stress-energy tensor~\eqref{stressE}.
In the diagonal case we consider, 
the unknowns are the six variables $(g^{ii}, k^i_{\ i})$ (no summation over $i$);
the Cauchy data is $(g^{ii}(0), k^i_{\ i}(0))$.

\section{Dynamical system formulation}
\label{dynamicalsystem}

In order to formulate Einstein equations with elastic
matter in Bianchi type I 
as a regular dynamical system we introduce alternative variables
and matter quantities.
Let
\begin{subequations}
\begin{align}
H & = -\frac{\mathrm{tr}\,k}{3}\:, \\
\label{Sigmaidef}
\Sigma_i & = -\frac{k^i_{\ i}}{H}-1 \quad \text{(no sum)} 
\qquad \qquad  \big( \: \Rightarrow \: \Sigma_1+\Sigma_2+\Sigma_3=0 \: \big) \:.
\end{align}
\end{subequations}
The Hubble scalar $H$ carries dimension, while the shear variables $\Sigma_1$, $\Sigma_2$, $\Sigma_3$
are dimensionless.
Evidently, the transformation between the variables $\big(k^1_{\ 1},k^2_{\ 2}, k^3_{\ 3}\big)$ 
and $(H, \Sigma_1,\Sigma_2, \Sigma_3)$, where $\Sigma_1 +\Sigma_2+\Sigma_3 = 0$,
is one-to-one.

In analogy to the new ``momentum variables'' we introduce new ``configuration
variables''. Let $(ijk)$ be a cyclic permutation of $(123)$. We define
\begin{subequations}
\begin{align}
G & = \mathrm{det}\, g^{-1} = g^{11} g^{22} g^{33} \:, \\
\label{yidef}
y_i & = 
\frac{g^{jj}}{g^{jj}+g^{kk}} \quad (\epsilon_{ijk} = +1) 
\qquad \qquad  \Big( \: \Rightarrow \: \frac{y_1}{1-y_1} \frac{y_2}{1-y_2} \frac{y_3}{1-y_3} = 1 \: \Big) \:.
\end{align}
\end{subequations}
The variable $G$ is dimensional, the variables $y_1$, $y_2$, $y_3$ are dimensionless;
by construction, we have $0< y_i < 1$ for all $i$.
The transformation of variables $(g^{11}, g^{22}, g^{33}) \mapsto (G, y_1,y_2,y_3)$,
where $(y_1,y_2,y_3)$ are subject to the constraint $(y_1 y_2 y_3)/[(1-y_1)(1-y_2)(1-y_3)] =1$,
is invertible, since
\begin{equation}\label{gii}
(g^{ii})^3 = G  \:\frac{1-y_j}{y_j} \frac{y_k}{1-y_k} \qquad(\epsilon_{ijk}=+1)\:.
\end{equation}

As a next step we normalize the matter quantities; we replace $(\rho, T^1_{\ 1}, T^2_{\ 2}, T^3_{\ 3})$
by $(\Omega, w_1, w_2, w_3)$ which we define as
\[
w_i=\frac{T^i_{\ i}}{\rho}\quad \text{(no sum)}\:, 
\qquad \qquad \qquad
\Omega=\frac{\rho}{3H^2}\:.
\]
It is customary to also introduce $w$ by
\[
w = \frac{1}{3} \sum_{i=1}^3 w_i = \frac{1}{3} \frac{\sum_{i} T^i_{\ i}}{\rho} = \frac{p}{\rho}\:.
\]
Since $p = a \rho$ for the elastic materials under consideration, see Section~\ref{elasticspacetimes}, 
we obtain 
\[
w = a\:.
\]

Expressed in terms of the new variables the 
(dimensionless) shear scalar $s^2$ is given by
\begin{equation}\label{s2iny}
s^2 = \frac{1}{12} \left[ \sum_{j=1}^3 \left(\frac{1-y_j}{y_j}+\frac{y_j}{1-y_j}\right)-6 \right]\:,
\end{equation}
and the quantities $w_i$ become
\begin{equation}\label{wi}
w_i=a+\frac{ab}{6}
\frac{\left(\frac{1-y_j}{y_j}-\frac{y_j}{1-y_j}\right)-\left(\frac{1-y_k}{y_k}-\frac{y_k}{1-y_k}\right)}%
{1+ ab\, s^2}\qquad\qquad(\epsilon_{ijk}=+1)\:,
\end{equation}
where we have used the elastic equations of state of Section~\ref{elasticspacetimes}.

Finally, we introduce a dimensionless time variable $\tau$ defined through
\begin{equation}\label{tandtau}
\partial_\tau=H^{-1}\partial_t\:,
\end{equation}
and we henceforth adopt the convention that a prime denotes differentiation with respect to $\tau$.

Written in the new dynamical variables
the Einstein equations split into the dimensional equations
\begin{equation}
\label{equationH}
H^\prime = -3 H\left[1-\frac{\Omega}{2} (1- w)\right]\:,
\qquad
G^\prime = - 6\, G 
\end{equation}
and a reduced set of dimensionless equations:
\begin{subequations}\label{dynamical}
\begin{align}
\label{Sigmaprime}
\Sigma_i^\prime & = -3\Omega\left[\frac{1}{2}(1-w)\Sigma_i-(w_i-w)\right] && (i =1,2,3)\\
\label{yprime}
y_i^\prime & = -2 y_i(1-y_i)\left[\Sigma_j-\Sigma_k\right]  \qquad(\epsilon_{ijk}=+1) &&  ( i =1,2,3)\:.
\end{align}
At the same time, the Hamiltonian constraint~\eqref{constraint} reads
\begin{equation}\label{constraint2}
1-\Sigma^2-\Omega=0\:,\qquad\text{where}\quad\Sigma^2:=\textfrac{1}{6}\sum_k\Sigma_k^2\:.
\end{equation}
\end{subequations}
We have thus arrived at the desired dynamical systems formulation of
our problem: The dynamical system~\eqref{dynamical} 
describes the dynamics of Bianchi type I elastic spacetimes,
where our choice of equations of state enters through the functions $w_i(y_1,y_2,y_3)$, $i=1, 2,3$.
Once the system~\eqref{dynamical} has been solved, the decoupled dimensional
equations~\eqref{equationH} can be integrated and the 
standard variables, in particular the
spatial metric, can be reconstructed. 

In addition to the dynamical system~\eqref{dynamical}, the following auxiliary 
equation for $\Omega$ will prove to be useful:
\begin{equation}\label{equationomega}
\Omega'=\Omega\left[3(1-w)\Sigma^2-\sum_k w_k\Sigma_k\right].
\end{equation}

In the remainder of this section we analyze in detail the state
space $\mathcal{X}$ of the dynamical system~\eqref{dynamical}.
The state space $\mathcal{X}$ is four-dimensional;
it is defined as the Cartesian product of two two-dimensional sets,
\begin{equation}
\mathcal{X} = \mathit{\Sigma} \times \mathcal{Y} \:,
\end{equation}
where
\begin{subequations}
\begin{align}
\mathit{\Sigma} & = \{ (\Sigma_1,\Sigma_2,\Sigma_3)\:\big|\: \sum_{i=1}^3 \Sigma_i = 0 \:\wedge\: \Sigma^2 < 1 \}\:, \\[1ex]
\mathcal{Y} & = \{ (y_1,y_2,y_3) \: \big|\: 0< y_i < 1 \;\forall i \:\:\wedge\:\: \prod_{i=1}^3 \frac{y_i}{1-y_i} = 1 \}\:.
\end{align}
\end{subequations}
The conditions on the variables are a direct consequence of the definitions~\eqref{Sigmaidef} 
and~\eqref{yidef} and the constraint~\eqref{constraint2}.

\begin{figure}[Ht]
\begin{center}
\subfigure[The Kasner disc $\mathit{\Sigma}$]{%
\psfrag{y1}[cc][cc][1][0]{$\Sigma_1$}
\psfrag{y2}[cc][cc][1][0]{$\Sigma_2$}
\psfrag{y3}[cc][cc][1][0]{$\Sigma_3$}
\psfrag{231}[lt][lt][1][0]{$\langle 231 \rangle$}
\psfrag{213}[tc][tc][1][0]{$\langle 213 \rangle$}
\psfrag{123}[rt][rt][1][0]{$\langle 123 \rangle$}
\psfrag{132}[rb][rb][1][0]{$\langle 132 \rangle$}
\psfrag{312}[bc][bc][1][0]{$\langle 312 \rangle$}
\psfrag{321}[lb][lb][1][0]{$\langle 321 \rangle$}
\psfrag{T1}[cc][cc][1][0]{$\mathrm{T}_1$}
\psfrag{T2}[cc][cc][1][0]{$\mathrm{T}_2$}
\psfrag{T3}[cc][cc][1][0]{$\mathrm{T}_3$}
\psfrag{Q1}[cc][cc][1][0]{$\mathrm{Q}_1$}
\psfrag{Q2}[cc][cc][1][0]{$\mathrm{Q}_2$}
\psfrag{Q3}[cc][cc][1][0]{$\mathrm{Q}_3$}
\includegraphics[width=0.45\textwidth]{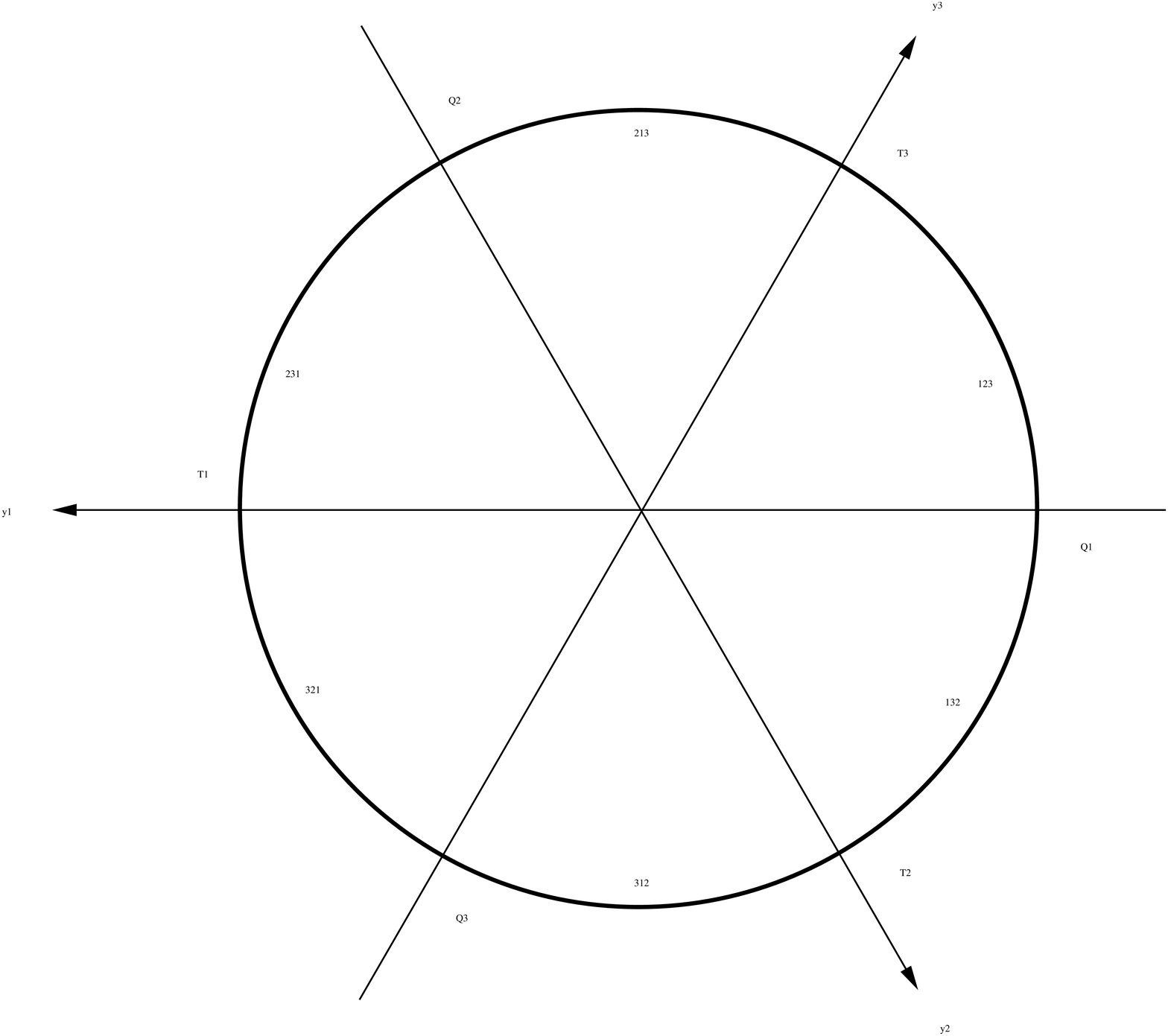}}
\qquad
\subfigure[The space $\mathcal{Y}$]{%
\psfrag{y1}[cc][cc][1][0]{$y_1$}
\psfrag{y2}[cc][cc][1][0]{$y_2$}
\psfrag{y3}[cc][cc][1][0]{$y_3$}
\psfrag{231}[lt][lt][1][0]{$[231]$}
\psfrag{a231}[bc][bc][1][60]{$(1,0,\rightarrow)$}
\psfrag{213}[tc][tc][1][0]{$[213]$}
\psfrag{a213}[bc][bc][1][0]{$(\leftarrow,0,1)$}
\psfrag{123}[rt][rt][1][0]{$[123]$}
\psfrag{a123}[bc][bc][1][-60]{$(0,\rightarrow,1)$}
\psfrag{132}[rb][rb][1][0]{$[132]$}
\psfrag{a132}[tc][tc][1][60]{$(0,1,\rightarrow)$}
\psfrag{312}[bc][bc][1][0]{$[312]$}
\psfrag{a312}[tc][tc][1][0]{$(\leftarrow,1,0)$}
\psfrag{321}[lb][lb][1][0]{$[321]$}
\psfrag{a321}[tc][tc][1][-60]{$(1,\rightarrow,0)$}
\psfrag{T1}[cc][cc][1][0]{$\mathcal{T}_1$}
\psfrag{T2}[cc][cc][1][0]{$\mathcal{T}_2$}
\psfrag{T3}[cc][cc][1][0]{$\mathcal{T}_3$}
\psfrag{Q1}[cc][cc][1][0]{$\mathcal{Q}_1$}
\psfrag{Q2}[cc][cc][1][0]{$\mathcal{Q}_2$}
\psfrag{Q3}[cc][cc][1][0]{$\mathcal{Q}_3$}
\includegraphics[width=0.45\textwidth]{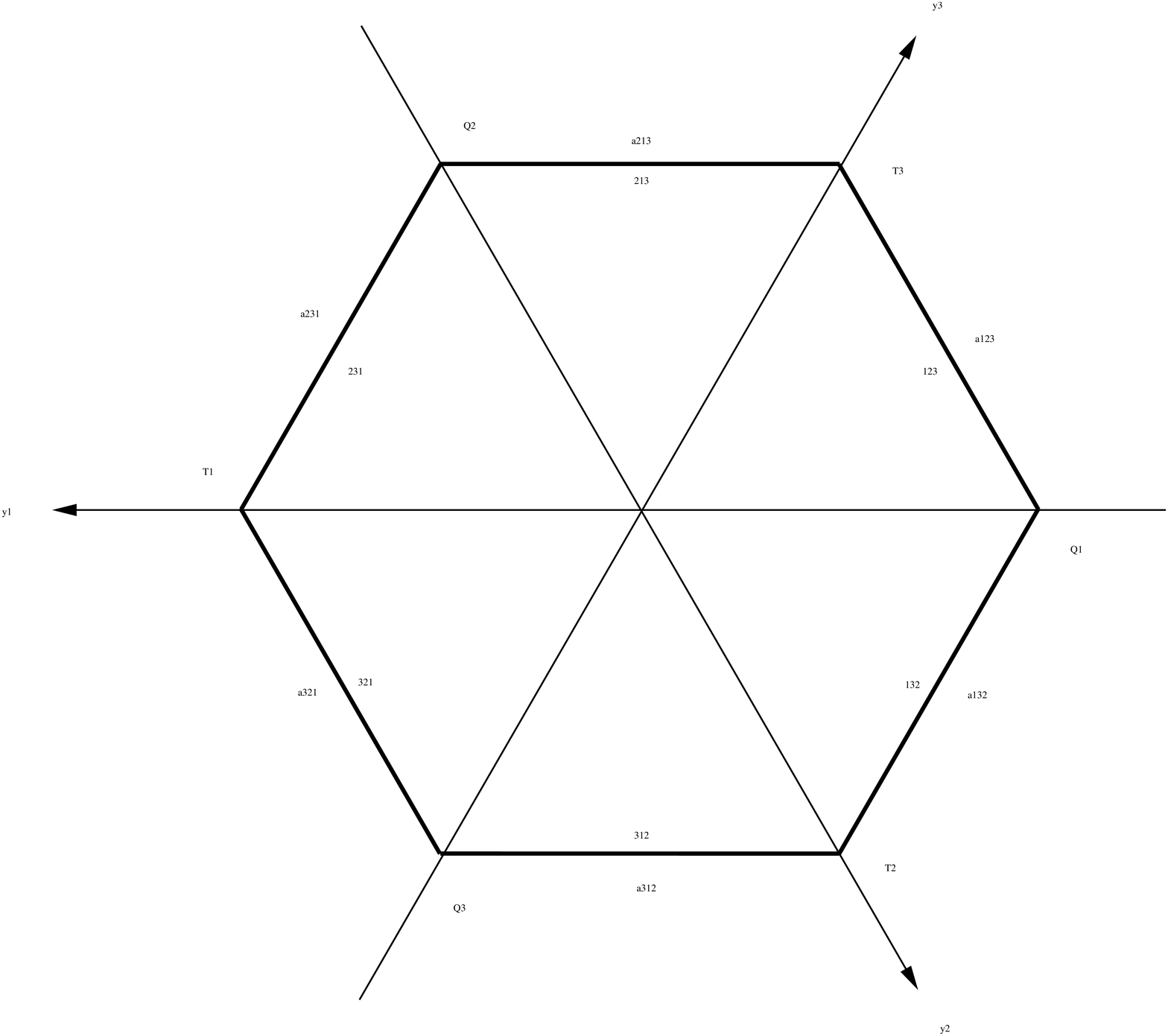}}
\caption{The four-dimensional state space $\mathcal{X}$ is the Cartesian product
of the Kasner disc $\mathit{\Sigma}$ and the surface $\mathcal{Y}$.
The space $\mathcal{Y}$ is most conveniently 
represented as (the interior of) a hexagon; the center of the hexagon is the point
$(y_1,y_2,y_3) = (1/2,1/2,1/2)$; for the edges the values of $(y_1,y_2,y_3)$ are 
given in the figure, where the arrows denote the directions
of increasing values (from $0$ to $1$).}
\label{SigmaandY}
\end{center}
\end{figure}

The set $\mathit{\Sigma}$ is the Kasner disc; it is usually 
depicted in a projection onto the plane with conormal $(1,1,1)$, see Figure~\ref{SigmaandY}.
The boundary of $\mathit{\Sigma}$ is the Kasner circle 
$\mathrm{KC} = \partial \mathit{\Sigma} = \{ \Sigma^2 =1 \}$.
The Kasner circle contains six special points, which are referred to as LRS points:
The three Taub points $\mathrm{T}_1$, $\mathrm{T}_2$, $\mathrm{T}_3$ 
given by $(\Sigma_1,\Sigma_2,\Sigma_3) = (2, -1,-1)$ and permutations,
and the three non-flat LRS points $\mathrm{Q}_1$, $\mathrm{Q}_2$, $\mathrm{Q}_3$ 
given by $(\Sigma_1,\Sigma_2,\Sigma_3) = (-2, 1,1)$ and permutations.
The six sectors of $\partial\mathit{\Sigma}$ are denoted by permutations of
the triple $\langle 123 \rangle$; by definition, $\Sigma_i < \Sigma_j <\Sigma_k$ holds
in sector $\langle ijk \rangle$.

The set $\mathcal{Y}$ is given as a two-dimensional surface in the interior of the 
unit cube $[0,1]^3$.
Its boundary $\partial\mathcal{Y}$ is the union of those six edges of the cube
that do not contain the vertices $(0,0,0)$ or $(1,1,1)$.
The projection of $\partial\mathcal{Y}$ onto an affine plane with conormal $(1,1,1)$
is a \textit{hexagon}, the surface $\mathcal{Y}$ itself its interior.
The center of the hexagon represents the point $(y_1,y_2,y_3) = (1/2,1/2,1/2)$.
In analogy to the Kasner circle, the six edges
of the hexagon $\partial\mathcal{Y}$ can be
regarded as six sectors, where sector $[ijk]$ is characterized
by the inequality $0 = y_i \leq y_j \leq y_k = 1$;
the six vertices of $\partial\mathcal{Y}$ separate the sectors from each other:
For $\mathcal{T}_i$ we have $(y_i,y_j,y_k) = (1,0,0)$,
for $\mathcal{Q}_i$ we have $(y_i,y_j,y_k) = (0,1,1)$;
see Figure~\ref{SigmaandY}.

A priori, by~\eqref{wi}, the quantities $w_i$ are given as smooth 
functions of $(y_1,y_2,y_3) \in \mathcal{Y}$ only.
It is a crucial fact, however, that these functions
admit a continuous extension to $\overline{\mathcal{Y}}$,
when we assume that $a b \neq 0$.
In this case it is straightforward to prove that
on sector $[ijk]$ of $\partial\mathcal{Y}$, $(w_i,w_j,w_k)$ is given by
\begin{equation}\label{wiboundary}
w_i = a + 2 \epsilon_{ijk} (1-y_j)\:,\qquad
w_j = a - 2 \epsilon_{ijk}\:, \qquad
w_k = a + 2 \epsilon_{ijk} y_j\:.
\end{equation}

It follows that the r.h.\ side
of the dynamical system~\eqref{dynamical} possesses an extension to 
the boundary of the state space, whereby we obtain a dynamical system 
on a compact state space $\overline{\mathcal{X}}$.
In particular, the analysis of the flow on the boundary $\partial \mathcal{X}$,
which is 
\begin{equation*}
\partial \mathcal{X} = \left(\partial \mathit{\Sigma} \times \overline{\mathcal{Y}} \right) 
\cup
\left(\mathit{\overline{\Sigma}} \times \partial \mathcal{Y} \right)\:,
\end{equation*}
will turn out to be essential for an understanding
of the global dynamics of the dynamical system.

We conclude this section with some remarks on energy conditions.
The dominant energy condition 
is expressed in the new matter variables as
\begin{equation}\label{dec2}
|w_i|\leqslant 1\qquad\forall\,i=1,2,3\:;
\end{equation}
the weak energy condition reads
\begin{equation}\label{weakenergycondition}
-1 \leqslant w_i  \qquad\forall\, i=1,2,3 \:;
\end{equation}
the strong energy condition is satisfied if~\eqref{weakenergycondition} holds and $a\geq -1/3$.
It is a simple consequence of~\eqref{wi} that
the dominant (and thus the weak) energy condition is satisfied
for perfect fluids, i.e., for $b =0$.
However, for elastic matter, when $|b| > 0$,
the dominant energy condition
is violated for $(y_1,y_2,y_3)\in\mathcal{Y}$ sufficiently close to the boundary $\partial\mathcal{Y}$. 
In fact, by~(\ref{wiboundary}), on each sector $[ijk]$ of $\partial\mathcal{Y}$ 
there is at least one quantity $w_i$ ($i=1,2,3$) 
such that $|w_i|>1$, and by continuity this inequality must hold in a neighborhood of that sector. 
(Note in this context that~\eqref{wiboundary} is independent of the value of $b \neq 0$.)
On the other hand, provided that $|a| < 1$, the dominant energy condition holds in some region of 
the interior of the state space, since $w_i(1/2,1/2,1/2)=a$ for all $i$ and thus 
by continuity $|w_i| < 1$ in a neighborhood of this point.


\section{Global dynamics}
\label{globaldyn}

The dynamical system~\eqref{dynamical} 
possesses one single equilibrium point
in the state space $\mathcal{X}$, which we call $\mathrm{F}$. 
This fixed point is given by 
\begin{subequations}
\begin{align}
& \mathrm{F}: \quad \Sigma_1=\Sigma_2=\Sigma_3=0 \:, \quad  y_1 = y_2 =y_3 =\textfrac{1}{2} \:;
\intertext{an alternative characterization is}
& \mathrm{F}: \quad \Sigma_1=\Sigma_2=\Sigma_3=0 \:, \quad  w_1=w_2=w_3= w = a\:, 
\end{align}
\end{subequations}
which is a direct consequence of~\eqref{wi} by taking into account the positivity of
the variables $y_i$. Since the principal pressures coincide, $p_1 = p_2 =p_3 = p$,
the fixed point $\mathrm{F}$ represents the flat isotropic FRW perfect fluid solution 
associated with the equation of state $p = a \rho$.

We now consider the function
\begin{equation}\label{Mdef}
M = (1-\Sigma^2)^{-1} \left(1 + ab\, s^2\right) \:,
\end{equation}
where $s^2$ is given by~\eqref{s2iny} and thus is a function of $(y_1,y_2,y_3)$.
Recall that we suppose $ab > 0$.
Accordingly, the function $M$ is positive on the state space $\mathcal{X}$; in fact,
$\min_{\mathcal{X}} M = 1$ and the minimum
$M = 1$ is attained at the fixed point $\mathrm{F}$ only.

A lengthy but straightforward computation, where we use 
the dynamical system~\eqref{dynamical}
and the functions $w_i(y_1,y_2,y_3)$, see~\eqref{wi}, 
leads to 
\begin{subequations}\label{M1properties}
\begin{equation}\label{M1prime}
M^\prime = -3 (1 - a) \Sigma^2 M\:.
\end{equation}
The computation of higher derivatives reveals that
\begin{equation}
M^{\prime\prime\prime} \Big|_{\Sigma^2 = 0} = -9 (1-a) M \sum_k (w_k-w)^2
\end{equation}
\end{subequations}
on the subset $\Sigma^2 =0$ of the state space.

Equation~\eqref{M1prime} suggests that
$a = 1$ is a special
case for our considerations, 
since then the r.h.\ side vanishes. 
We thus distinguish two cases: $a < 1$ and $a=1$. 

\subsection*{The case $\bm{a < 1}$}

When we consider an elastic equation of state with $a <  1$ (and $a b > 0$), 
Equation~\eqref{M1properties} implies that
$M^\prime < 0$ when $\Sigma^2 \neq 0$
and $M^{\prime\prime\prime}|_{\Sigma^2 = 0} < 0$ except at the point $\mathrm{F}$;
therefore,
$M$ is a strictly monotonically decreasing function 
on $\mathcal{X} \backslash \mathrm{F}$.
The existence of a monotone function allows us to prove two
central theorems.

\begin{Theorem}[\textbf{Future asymptotics}]\label{futurethm}
All orbits in the state space $\mathcal{X}$ converge to the fixed point 
$\mathrm{F}$ when $\tau\rightarrow +\infty$.
\end{Theorem}

\textit{Interpretation of the theorem}.
The fixed point $\mathrm{F}$ corresponds to a FRW perfect fluid solution
associated with the equation of state $p = a \rho$;
the theorem thus states that each Bianchi type I model with elastic matter
obeying an equation of state with $a\in [-1,1)$ and $ab > 0$
isotropizes toward the future and behaves like an (infinitely
diluted) isotropic perfect fluid solution in the asymptotic regime.

\begin{proof}
The function $M $ is strictly monotonically decreasing along every orbit
in the invariant set $\mathcal{X} \backslash \mathrm{F}$. 
The monotonicity principle~\cite{FHU, WE} implies that the $\omega$-limit of every orbit
must be contained on the boundary, which is $\partial \mathcal{X} \cup \mathrm{F}$.
Since $M  = +\infty$ on $\partial \mathcal{X}$, the boundary $\partial \mathcal{X}$
is excluded, which leaves the fixed point
$\mathrm{F}$ as the only possible $\omega$-limit.
\end{proof}

\begin{Theorem}[\textbf{Past asymptotics}]\label{pastthm}
The $\alpha$-limit of every orbit in $\mathcal{X}\backslash \mathrm{F}$ 
is a subset of the boundary $\partial \mathcal{X}$ of the state space.
\end{Theorem}

\begin{proof}
The monotonicity principle implies that the $\alpha$-limit 
must be contained on $\partial \mathcal{X} \cup \mathrm{F}$.
The point $\mathrm{F}$ is excluded, however, since
$M = \min_{\mathcal{X}} M = 1$ at $\mathrm{F}$.
\end{proof}

Theorem~\ref{pastthm} describes the behavior of Bianchi type I models 
toward the initial singularity (which we choose to be $t = 0$).
To see this we simply note that the inequalities
$-3 H \leq H^\prime \leq -3/2 (1+ a) H$ follow from~\eqref{equationH},
so that~\eqref{tandtau} can be integrated to yield
a positive function $t(\tau)$ that satisfies $t\rightarrow 0$ as $\tau\rightarrow -\infty$.
(In this context it is necessary to assume $a>-1$; 
the case $a = -1$ requires a different argument that
involves more detailed information on the $\alpha$-limits of orbits in $\mathcal{X}$, 
see Section~\ref{past}.)

Since the boundary $\partial\mathcal{X}$ contains the Kasner circle(s), see
Section~\ref{past}, the theorem suggests that the Kasner solutions
will play an essential role in the past asymptotic dynamics
of Bianchi type I solutions with elastic matter.
However, in Section~\ref{past} we will prove that
there does not exist any solution that converges to a Kasner solution
as $t\rightarrow 0$; instead we observe oscillatory behavior
toward the singularity.
In order to derive these results concerning the past asymptotic dynamics of solutions 
we must analyze the flow of the dynamical system on the boundary $\partial\mathcal{X}$,
which will be done in Section~\ref{past}.

\subsection*{The case $\bm{a =1}$}

When $a =1$, the function $M$ 
is \textit{constant} along the orbits
of the dynamical system, i.e., 
for every $\mathbb{R} \ni m > 1$, the hypersurface \[
\mathcal{M}_m = \left\{ (\Sigma_1,\Sigma_2,\Sigma_3, y_1,y_2,y_3) \in \mathcal{X}\:|\: M= m >1\right\}
\]
is an invariant subset in $\mathcal{X}$.
In other words, $M$ acts as a conserved ``energy''; solutions with
the ``energy'' $M = m$ are contained on $\mathcal{M}_m$. 
When $m=1$, we have $\mathcal{M}_1 = \{F\}$.

It is not difficult to show from~\eqref{s2iny} 
that each surface $s^2 = \mathrm{const}$ is
a (topological) sphere in the unit cube $(0,1)^3$, which is 
centered at $(y_1,y_2,y_3) = (1/2,1/2,1/2)$.
Accordingly, $s^2 = \mathrm{const}$ defines a
closed curve in $\mathcal{Y}$.
Consequently, when we rewrite~\eqref{Mdef} in the form
\begin{equation}\label{S3hyper}
M \Sigma^2 + ab\, s^2 = M-1 \qquad (M = m >1) \:,
\end{equation}
we conclude that $\mathcal{M}_m$
represents a topological $S^3$ hypersphere 
in $\mathcal{X}$ whose
center is the fixed point $\mathrm{F}$.
In particular, $\mathcal{M}_m$ does not intersect $\partial \mathcal{X}$. 
We have proved the following result.

\begin{Theorem}\label{caseaeq1}
For the $\omega$-limit set $\omega(\gamma)$ of an orbit $\gamma$ in $\mathcal{X}\backslash \mathrm{F}$ 
we have: $\mathrm{F}\notin\omega(\gamma)$ and $\omega(\gamma)\cap\partial\mathcal{X} = \emptyset$. 
An identical statement holds for the $\alpha$-limit set of $\gamma$. 
\end{Theorem}

\textit{Interpretation of the theorem}.
If the equation of state of the elastic matter is such that $a=1$, then the associated
Bianchi type I solutions of the Einstein equations
do not isotropize toward the future (and neither toward the singularity).
Furthermore, the solution cannot be approximated by Kasner solutions
at any time and neither asymptotically 
(since Kasner solutions are represented by points on $\partial\mathcal{X}$, see Section~\ref{past}).

From~\eqref{S3hyper} it follows that the maximum shear of a solution with ``energy''
$M = m$ is given by $s^2 = (m-1)/(a b)$, while
$\Sigma^2$ remains bounded by $\Sigma^2 \leq 1-1/m$, so that $\Omega \geq 1/m$.

\section{LRS solutions}
\label{LRSsolutions}

In this section we consider a special case of Bianchi type I models:
Locally rotationally symmetric (LRS) models. 
The additional symmetry that is imposed reduces 
the number of degrees of freedom, which facilitates the analysis of the
(past asymptotic) dynamics of solutions in both cases $a<1$ and $a=1$.

A Bianchi type I solution of the Einstein equations 
is locally rotationally symmetric (LRS) if 
$\Sigma_j \equiv \Sigma_k$ and $w_j \equiv  w_k$
for some pair $j\neq k$.
Let $(ijk)$ be the completion of the pair $(j,k)$ to a permutation of the triple $(123)$.
The first condition, i.e., $\Sigma_j \equiv \Sigma_k$, 
implies that $y_i \equiv \mathrm{const}$ via~\eqref{yprime}, 
and thus $g^{jj}/g^{kk} \equiv \mathrm{const}$ according to~\eqref{yidef};
by a possible rescaling of the spatial coordinates we obtain LRS geometry, i.e., $g^{jj} \equiv g^{kk}$.
The second condition, i.e., $w_j \equiv w_k$, states that
the matter content is compatible with LRS symmetry; it guarantees that $y_i \equiv 1/2$
(so that $g^{jj} \equiv g^{kk}$ automatically).
To see this we use~\eqref{wi} and observe that $w_j = w_k$ iff
\begin{equation*}
\frac{1-y_j}{y_j} - \frac{y_j}{1-y_j} + \frac{1-y_k}{y_k} - \frac{y_k}{1-y_k} =
2 \left( \frac{1-y_i}{y_i} - \frac{y_i}{1-y_i}\right) \:;
\end{equation*}
multiplication with $(y_j y_k)/[(1-y_j)(1-y_k)]$, where we use the constraint~\eqref{yidef}
on the variables $(y_i,y_j,y_k)$, yields
\begin{equation*}
\left(\frac{y_j}{1-y_j} +\frac{y_k}{1-y_k}\right) \left(1 -\frac{y_j}{1-y_j}\frac{y_k}{1-y_k} \right)
= - 2 \left(1 - \frac{y_j}{1-y_j}\frac{y_k}{1-y_k}\right)\left(1 + \frac{y_j}{1-y_j}\frac{y_k}{1-y_k}\right)
\end{equation*}
and hence
\begin{equation*}
w_j = w_k \:\Leftrightarrow \: \left(1 -\frac{y_j}{1-y_j}\frac{y_k}{1-y_k} \right) = 0 \: \Leftrightarrow \:
y_j = 1 -y_k \:\Leftrightarrow \: y_i = \frac{1}{2}\:.
\end{equation*}

\begin{figure}[Ht]
\begin{center}
\subfigure[$\mathrm{LRS}_i \cap \mathit{\Sigma}$]{%
\psfrag{y1}[cc][cc][1][0]{$\Sigma_1$}
\psfrag{y2}[cc][cc][1][0]{$\Sigma_2$}
\psfrag{y3}[cc][cc][1][0]{$\Sigma_3$}
\psfrag{l1}[cc][cc][1][0]{$\mathrm{LRS}_1$}
\psfrag{l2}[rc][cc][1][0]{$\mathrm{LRS}_2$}
\psfrag{l3}[lc][cc][1][0]{$\mathrm{LRS}_3$}
\includegraphics[width=0.45\textwidth]{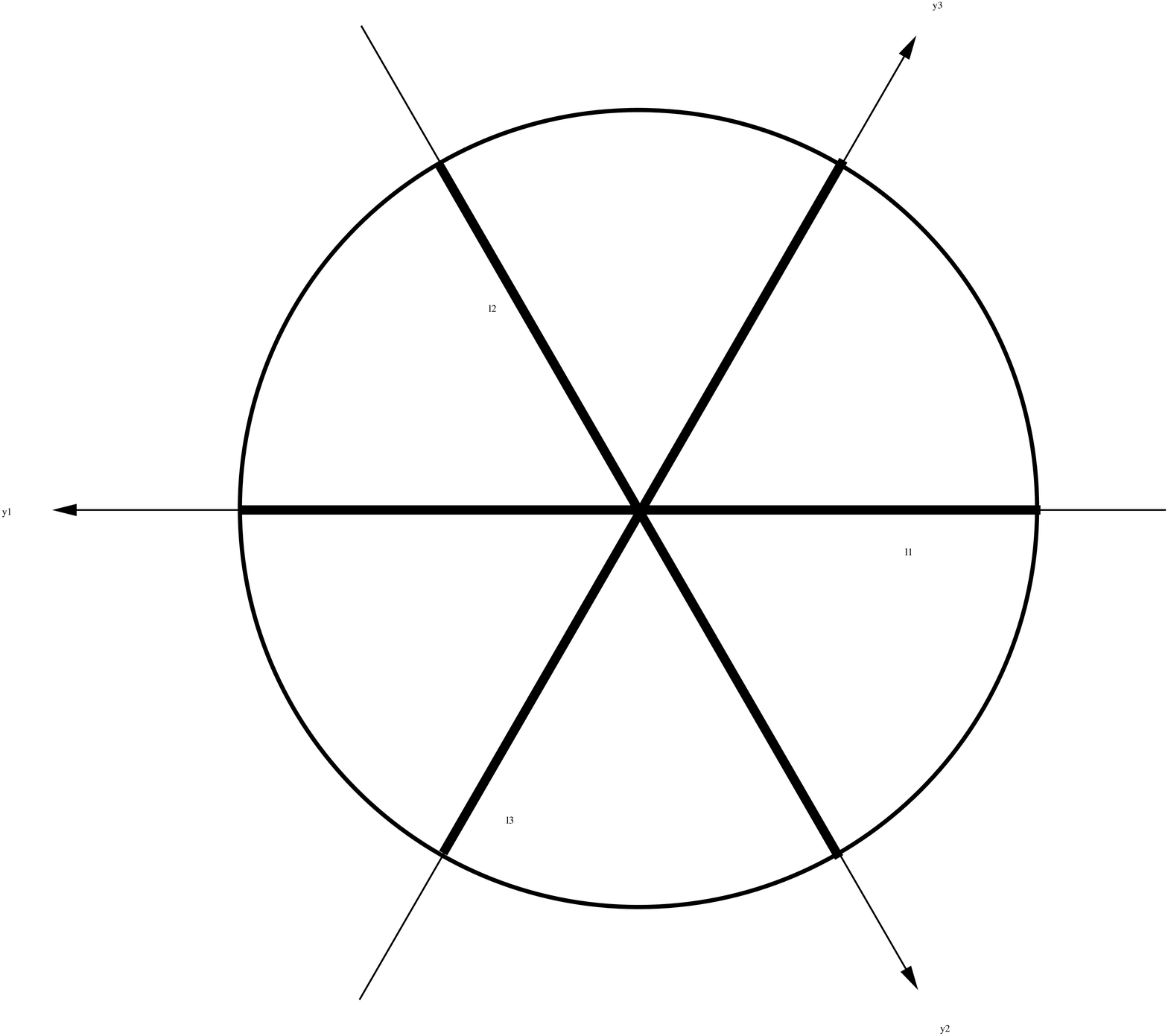}}
\qquad
\subfigure[$\mathrm{LRS}_i \cap \mathcal{Y}$]{%
\psfrag{y1}[cc][cc][1][0]{$y_1$}
\psfrag{y2}[cc][cc][1][0]{$y_2$}
\psfrag{y3}[cc][cc][1][0]{$y_3$}
\psfrag{l1}[cc][cc][1][0]{$\mathrm{LRS}_1$}
\psfrag{l2}[rc][cc][1][0]{$\mathrm{LRS}_2$}
\psfrag{l3}[lc][cc][1][0]{$\mathrm{LRS}_3$}
\includegraphics[width=0.45\textwidth]{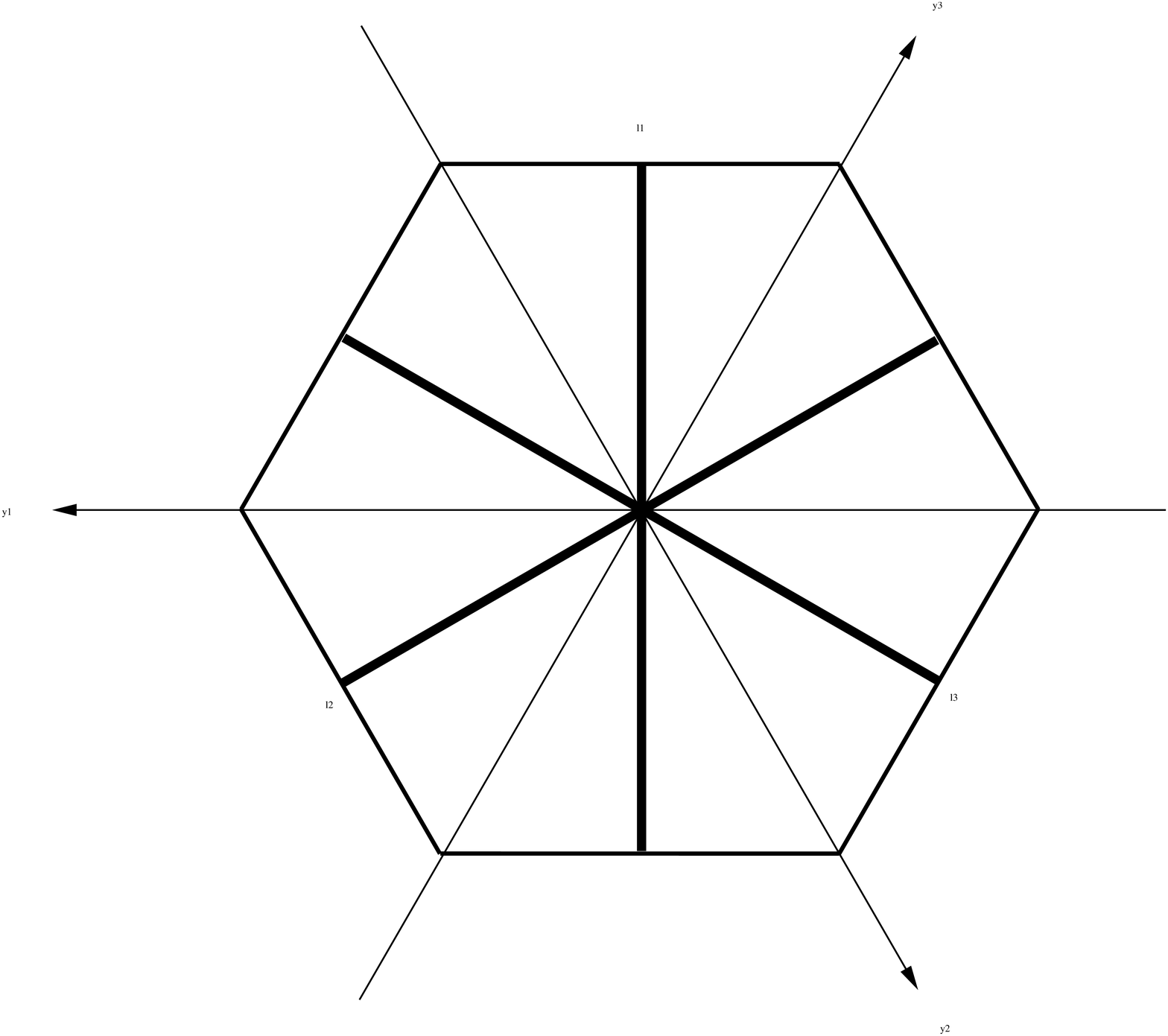}}
\caption{In the state space $\mathcal{X} = \mathit{\Sigma} \times \mathcal{Y}$ 
there exist three invariant subsets that 
can be identified as LRS subsets.
We depict the intersection of $\mathrm{LRS}_1$, $\mathrm{LRS}_2$, $\mathrm{LRS}_3$ 
with the factors $\mathit{\Sigma}$ and $\mathcal{Y}$ of the state space.}
\label{LRSsubsets}
\end{center}
\end{figure}

In the state space $\mathcal{X}$, 
the conditions $\Sigma_j = \Sigma_k$
and $y_i = \textfrac{1}{2}$ ($\Leftrightarrow w_j = w_k$)
define three invariant subsets which we denote by $\mathrm{LRS}_i$, $i=1,2,3$,
see Figure~\ref{LRSsubsets}.
In the dynamical systems representation of
Bianchi type~I elastic cosmologies,
LRS configurations are given by
orbits on one of these invariant subsets.
In the following we thus analyze the dynamics of LRS solutions by studying
the flow of the dynamical system on the LRS subsets $\mathrm{LRS}_i$.

Let again $(ijk)$ be a cyclic permutation of $(123)$.
Consider the subset $\mathrm{LRS}_i \subset \mathcal{X}$, which is given by the
conditions
\begin{equation*}
y_i = \textfrac{1}{2} \quad \big( \Leftrightarrow\, w_j = w_k \,\Leftrightarrow\, y_j = 1- y_k \big)
\quad \text{and} \quad
\Sigma_j = \Sigma_k \quad \big( \Leftrightarrow\: \Sigma_i = -2 \Sigma_j = -2 \Sigma_k \big)\:.
\end{equation*}
On $\mathrm{LRS}_i$, the shear scalar $s^2$ reads
\begin{equation*}
s^2 = \frac{1}{6} \: \frac{(1-2 y_j)^2}{y_j (1-y_j)} \,=\, \frac{1}{6} \: \frac{(1-2 y_k)^2}{y_k (1-y_k)}
\end{equation*}
and the matter quantities become
\begin{equation}\label{wji}
w_j = a - \frac{ab}{6}\:\frac{1- 2 y_j}{y_j(1-y_j) +\frac{ab}{6} (1-2 y_j)^2} = w_k \:,
\qquad 
w_i = 3a - 2 w_j\:.
\end{equation}

The dynamical system~\eqref{dynamical} reduces to
\begin{subequations}\label{dynamicalLRS}
\begin{align}
y_j' & =-6 y_j (1-y_j)\Sigma_j\:, \\
\Sigma_j' & = -3 (1-\Sigma_j^2) \left[\frac{1}{2}(1-a)\Sigma_j-(w_j-a)\right],
\end{align}
\end{subequations}
where we have used that $\Omega = 1- \Sigma_j^2$ under the present assumptions.

The state space $\mathrm{LRS}_i$ can be represented as (the
interior of) the rectangle $(-1,1) \times (0,1) \ni (\Sigma_j,y_j)$. 
Since $w_j$ extends smoothly
to $y_j =0$ and $y_j = 1$, the dynamical system~\eqref{dynamicalLRS}
extends smoothly to the compact space $[-1,1]\times [0,1]$.
The four sides of the rectangle are invariant subspaces; when we
exclude the vertices from our considerations, we find:
\begin{alignat*}{5}
& \bullet \quad\Sigma_j & & = \pm 1 & \quad \Rightarrow \quad &  & y_j^\prime  & = \mp \: 6\, y_j (1 - y_j) & & \lessgtr\, 0  \\[0.5ex]
& \bullet \quad y_j & & = \:0 & \quad \Rightarrow \quad & & \Sigma_j^\prime  & = -\textfrac{3}{2} \Omega \left[ (1-a)\Sigma_j +2 \right] & & <\, 0\\[0.5ex]
& \bullet \quad y_j & & = \:1 & \quad \Rightarrow \quad & & \Sigma_j^\prime  & = -\textfrac{3}{2} \Omega \left[ (1-a)\Sigma_j -2 \right] & &> \,0
\end{alignat*}
The four vertices of the rectangle are fixed points. (The notation is chosen to agree
with the conventions of Section~\ref{past}.)
\begin{center}
\begin{tabular}{|c|c|c|c|}\hline
Fixed point & $(\Sigma_i,\Sigma_j, \Sigma_k)$ & $(y_i,y_j,y_k)$  & Fixed point represents\\ \hline
$\mathrm{Q}_{(j\bm{i}k)}$ &  $ \mathrm{Q}_i = (-2,+1,+1)$ &  $(1/2, 0, 1)$ & non-flat LRS Kasner solution\\
$\mathrm{Q}_{(k\bm{i}j)}$ &  $ \mathrm{Q}_i = (-2,+1,+1)$ &  $(1/2, 1, 0)$ & non-flat LRS Kasner solution\\
$\mathrm{T}_{(j\bm{i}k)}$ &  $ \mathrm{T}_i = (+2,-1,-1)$ &  $(1/2, 0, 1)$ & Taub solution (flat LRS Kasner)\\
$\mathrm{T}_{(k\bm{i}j)}$ &  $ \mathrm{T}_i = (+2,-1,-1)$ &  $(1/2, 1, 0)$ & Taub solution (flat LRS Kasner)\\\hline
\end{tabular}
\end{center}

From this analysis it follows that the boundary of the state space $\mathrm{LRS}_i$
forms a heteroclinic cycle:
\begin{equation}\label{heterocycle}
\begin{CD}
\mathrm{T}_{(k\bm{i}j)} @>>> \mathrm{Q}_{(k\bm{i}j)} \\
@AAA @VVV \\
\mathrm{T}_{(j\bm{i}k)} @<<< \mathrm{Q}_{(j\bm{i}k)} 
\end{CD}
\end{equation}

In the interior of the state space $\mathrm{LRS}_i$ 
there exists one single fixed point: The FRW perfect fluid
fixed point $\mathrm{F}$; recall that $\Sigma_i = \Sigma_j =\Sigma_k =0$
and $y_i = y_j =y_k =1/2$ at $\mathrm{F}$.

\begin{figure}[Ht]
\begin{center}
\psfrag{yj}[cc][cc][1][0]{$y_j$}
\psfrag{sigj}[cc][cc][1][0]{$\Sigma_j$}
\psfrag{Qkij}[cc][cc][1][0]{$\text{Q}_{(k\bm{i}j)}$}
\psfrag{Qjik}[cc][cc][1][0]{$\text{Q}_{(j\bm{i}k)}$}
\psfrag{Tjik}[cc][cc][1][0]{$\text{T}_{(j\bm{i}k)}$}
\psfrag{Tkij}[cc][cc][1][0]{$\text{T}_{(k\bm{i}j)}$}
\psfrag{F}[cc][cc][1][0]{F}
\includegraphics[width=1\textwidth]{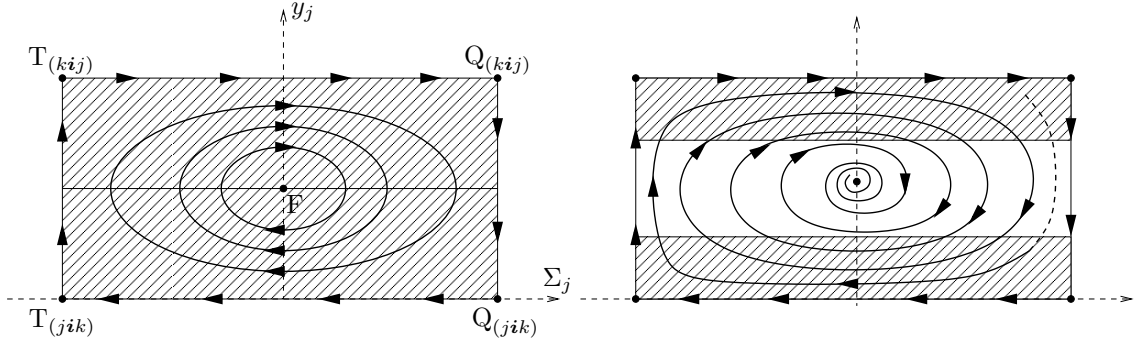}
\caption{Phase portraits of LRS solutions for $a=1$ (left) and $a\in [-1,1)$ (right). 
The dominant energy condition is violated in the shadowed region 
(for $a=-1$, this is everywhere except at $y_j=1/2$).
It is assumed that $ab\leq 3$ and, for $a\in [-1,1)$, 
that $ab>\frac{3}{32}(1-a)^2$ (oscillatory approach toward F).}
\label{lrspicture}
\end{center}
\end{figure}

To analyze the global dynamics on $\mathrm{LRS}_i$ we distinguish
the cases $a = 1$ and $a\neq 1$ as in Section~\ref{globaldyn}.
First, let $a =1$.
The considerations of Section~\ref{globaldyn} imply that
there exists a family of invariant subsets
$\{\mathcal{M}_m\: |\: m >1\}$. 
For each $m$, the surface $\mathcal{M}_m$ is a three-dimensional 
hypersphere and its intersection with the two-dimensional surface $\mathrm{LRS}_i$
yields a closed curve with center $\mathrm{F}$.
By construction, this closed curve is an orbit of the dynamical system on $\mathrm{LRS}_i$.
More explicitly, we see that
\begin{equation}
M = (1 - \Sigma^2)^{-1} (1 +ab\, s^2) = 
(1 -\Sigma_j^2)^{-1} 
\left(1 + \frac{ab}{6} \: \frac{(1-2 y_j)^2}{y_j (1-y_j)}\right) = m > 1
\end{equation}
defines a family of periodic orbits in $\mathrm{LRS}_i$ which are
centered at the fixed point $\mathrm{F}$.
The phase portrait of
the dynamical system in the case $a =1$ 
is represented in Figure~\ref{lrspicture}.
(Note that in the limiting fluid case, i.e., $a =1$ with $b = 0$,
the orbits are no longer periodic but straight lines $\Sigma_j = \mathrm{const}$.)

Second we consider elastic matter with the property $a\neq 1$.
The global dynamics on $\mathrm{LRS}_i$ follows
from 
Theorem~\ref{futurethm} and Theorem~\ref{pastthm}
in conjunction with the analysis of the boundary of $\mathrm{LRS}_i$:

\begin{Corollary}
The $\omega$-limit of
every orbit on $\mathrm{LRS}_i$ is the fixed point $\mathrm{F}$, the $\alpha$-limit
is the heteroclinic cycle~\eqref{heterocycle}.
\end{Corollary}

\textbf{Interpretation of the corollary}.
Each LRS Bianchi type I model with elastic matter 
obeying an equation of state with $a\in [-1, 1)$ and $ab > 0$
isotropizes toward the future and behaves like an (infinitely
diluted) isotropic perfect fluid solution in the asymptotic regime.
Toward the singularity we observe \textit{oscillatory behavior}
between the non-flat LRS Kasner solution and the Taub solution.
This is in stark contrast to the behavior of 
perfect fluid solutions, which converge to either the
non-flat LRS Kasner solution or to the Taub solution
as $t\rightarrow 0$.

The past asymptotic dynamics of solutions as described by the corollary 
is intimately connected with the violation of energy conditions.
As a matter of course, the general statement of Section~\ref{dynamicalsystem} 
also applies in the LRS case: 
While the dominant energy condition (and thus the weak energy condition) is
satisfied in a neighborhood of the fixed point $\mathrm{F}$ 
(where it is assumed that $|a| < 1$), 
we observe energy condition violation in a neighborhood of the boundaries
$y_j =0$ and $y_j =1$.

The discussion of the issue of energy condition violation 
is facilitated when we assume the upper bound $a b\leq 3$. 
Under this condition it is straightforward to show that $(w_j-a)$ is a monotonically
increasing function of $y_j\in [0,1]$ with range $[-1,1]$; accordingly, $(w_i-a)$ is decreasing
with range $[-2,2]$. A particularly simple case, which 
displays all the relevant features of the general case $a b\leq 3$,
is $a b = 3/2$, since $(w_i, w_j, w_k)$ become linear in $y_j$, i.e.,
\begin{equation}\label{wispecial}
w_j = a + 2 y_j -1 = w_k \:, \qquad\quad w_i = a + 2 - 4 y_j \:.
\end{equation}
A straightforward calculation yields the following results for this special case:
\begin{itemize}
\item The weak energy condition is satisfied iff
\[ 
y_j\in \left[-\frac{a}{2},\frac{a+3}{4}\right]\,.
\]
This interval collapses to the point $y_j = 1/2$ when $a=-1$.
\item The dominant energy condition is satisfied iff
\[
y_j\in \left[-\frac{a}{2},1-\frac{a}{2}\right]\cap\left[\frac{a+1}{4},\frac{a+3}{4}\right]\,;
\]
see Figure~\ref{lrspicture}.
This domain reduces to the point $y_j =1/2$ when $a = \pm1$.
\end{itemize}
A general property that is worth observing is the presence of 
regions where the dominant energy condition is violated, while 
the weak condition is satisfied. 

For a given $a\in (-1,1)$, the regions of energy condition violation
become smaller when we let $|b|\rightarrow 0$.
However, the general statement of Section~\ref{dynamicalsystem} 
applies for all $b$, no matter how small: 
There exists a neighborhood of the boundaries $y_j =0$ and $y_j =1$, 
where the energy conditions are violated.
To first order in $b$, the region where the energy conditions hold
is characterized by 
\begin{equation*}
\max \left\{ \frac{(1+a) a b}{6(1-a)},-\frac{a^2 b}{6(1+a)}\right\}
\leq\, y_j \, \leq
\min  \left\{ 1- \frac{(1-a) a b}{6(1+a)},1 -\frac{a^2 b}{6(1-a)}\right\}\:;
\end{equation*}
only when $b = 0$, i.e., in the fluid case, we obtain $0\leq y_j \leq 1$.

To conclude this section we study in detail the future
asymptotics of solutions, which in the state space description of
the dynamics corresponds to investigating the flow in
the neighborhood of the fixed point $\mathrm{F}$.

Since the r.h.\ side of the dynamical system is smooth in
a neighborhood of $\mathrm{F}$ we can perform 
a local dynamical systems analysis.
The linearization of the dynamical system at $\mathrm{F}$
possesses the eigenvalues
\begin{equation*}
\lambda_1 = \frac{3}{4} \left( -(1-a) -\sqrt{(1-a)^2 - \frac{32}{3}\: ab}\right)\:,
\qquad
\lambda_2 = \frac{3}{4} \left( -(1-a) +\sqrt{(1-a)^2 - \frac{32}{3}\: ab}\right)\:.
\end{equation*}
The eigenvectors associated with $\lambda_{1,2}$ are
\begin{equation}
v_{1,2} = \left(a b\,,\, \frac{3}{16} \,
\left[(1-a) \mp \sqrt{(1-a)^2 -\frac{32}{3}\:a b} \right]\right)^{T}\:.
\end{equation}
It is immediate that
\begin{itemize}
\item the eigenvalue(s) 
are real (and negative), 
if $ab \leqslant \frac{3}{32}(1-a)^2$; in this case
$\mathrm{F}$ is a stable node;
\item 
the eigenvalues are complex (with negative real part), 
if $ab > \frac{3}{32}(1-a)^2$; in this case
the fixed point $\mathrm{F}$ is a stable focus
and the solutions' approach to $\mathrm{F}$ as $\tau\rightarrow \infty$
is oscillatory; see Figure~\ref{lrspicture}.
\end{itemize}
The late time behavior of Bianchi type I models with elastic matter 
is thus characterized by
\begin{itemize}
\item monotonic isotropization if $a b \leqslant (3/32)(1-a)^2$;
\item oscillatory isotropization if $ab > (3/32)(1-a)^2$;
both the amplitude of the oscillations and the frequency
are decreasing as $t\rightarrow \infty$.
\end{itemize}

\section{Anti-LRS solutions}
\label{antiLRSsolutions}

The LRS subsets of the Kasner disc $\overline{\mathit{\Sigma}}$ are defined
by requiring that $\Sigma_j = \Sigma_k$ for some pair $(j,k)$. 
Analogously, we define the three \textit{anti-LRS} subsets 
by setting $\Sigma_j = -\Sigma_k$ for
some pair $(j,k)$; consequently, $\Sigma_i = 0$, where $(ijk)$ denotes 
the completion of the pair $(j,k)$ to a
permutation of the triple $(123)$. 
Let $\epsilon_{ijk} = 1$;
the six anti-LRS points on the Kasner circle $\partial \mathit{\Sigma}$
are given by the three points 
$(\Sigma_i,\Sigma_j,\Sigma_k) = (-\sqrt{3},0,\sqrt{3})\in \langle ijk \rangle$,
which we denote by $\mathrm{P}_j$, $j=1,2,3$,
and by the three points 
$(\Sigma_i,\Sigma_j,\Sigma_k) = (\sqrt{3},0,-\sqrt{3}) \in \langle kji \rangle$,
which we denote by $\mathrm{S}_j$, $j=1,2,3$.
(The notation is chosen to complement the standard notation $\mathrm{Q}_j$,
$\mathrm{T}_j$ for the LRS points.)
Since the angular distance on the Kasner circle is a na\"{i}ve measure for
the difference between Kasner states,
anti-LRS states are those which are maximally different from LRS states;
hence the name.

Let again $(ijk)$ be a cyclic permutation of $(123)$. 
We define the anti-LRS subset ${}^{\mathrm{a}}\mathrm{LRS}_i$ in the state space $\mathcal{X}$
by
\begin{subequations}
\begin{alignat}{3}
& {}^{\mathrm{a}}\mathrm{LRS}_i: \qquad &  & \Sigma_j = -\Sigma_k \quad & & \wedge\quad  y_j = y_k \:.
\intertext{A equivalent definition is}
& {}^{\mathrm{a}}\mathrm{LRS}_i: \qquad &  & \Sigma_i = 0  \quad & & \wedge\quad  w_i = w = a \: ;
\end{alignat}
\end{subequations}
here, $i=1,2,3$, so that there exist three anti-LRS subsets: ${}^{\mathrm{a}}\mathrm{LRS}_i$, $i=1,2,3$.
Since $\Sigma_i^\prime = 0$ and $(y_j - y_k)^\prime = 0$ on ${}^{\mathrm{a}}\mathrm{LRS}_i$,
cf.~\eqref{dynamical},
these subsets are \textit{invariant subsets} in $\mathcal{X}$.
Orbits on ${}^{\mathrm{a}}\mathrm{LRS}_i$ generate anti-LRS solutions of Bianchi type I;
the metric of these models is characterized by the condition 
that $g_{ii}$ be the geometric mean of $g_{jj}$ and $g_{kk}$, i.e.,
\begin{equation}
g_{ii} = \sqrt{\, g_{jj}\, g_{kk}\, } \:.
\end{equation}
Evidently, a solution is both LRS and anti-LRS iff it is isotropic.

On the anti-LRS subset ${}^{\mathrm{a}}\mathrm{LRS}_i$ the dynamical system~\eqref{dynamical}
reduces to
\begin{subequations}\label{dynamicalantiLRS}
\begin{align}
y_j' & = 2 y_j (1-y_j)\Sigma_j\:, \\
\Sigma_j' & = -3 (1-\Sigma_j^2/3) \left[\frac{1}{2}(1-a)\Sigma_j-(w_j-a)\right]\:,
\end{align}
\end{subequations}
where we have used that $\Omega = 1 - \Sigma_j^2/3$.
It is not difficult to show that $w_j$ reads
\begin{equation}
w_j = a + \frac{2 a b (1 - 2 y_j) [1 -y_j (1-y_j) ]}{12 y_j^2 (1-y_j)^2 + a b (1 -2 y_j)^2 [1 - 2 y_j (1-y_j)]}\:.
\end{equation}
The (closure of the) state space ${}^{\mathrm{a}}\mathrm{LRS}_i$ can be represented as the rectangle
$[-\sqrt{3},\sqrt{3}] \times [0,1] \ni (\Sigma_j, y_j)$.
The four vertices are fixed points.
\begin{center}
\begin{tabular}{|c|c|c|c|}\hline
Fixed point & $(\Sigma_i,\Sigma_j, \Sigma_k)$ & $(y_i,y_j,y_k)$  & Interpretation\\ \hline
$\mathrm{P}_{\mathcal{T}_i} = \{\mathrm{P}_i\} \times \{\mathrm{T}_i\} $ &  
$ \mathrm{P}_i = (0,+\sqrt{3},-\sqrt{3})$ &  $\mathcal{T}_i = (1, 0, 0)$ & anti-LRS Kasner solution\\
$\mathrm{P}_{\mathcal{Q}_i} = \{\mathrm{P}_i\} \times \{\mathrm{Q}_i\}$ &  
$ \mathrm{P}_i = (0,+\sqrt{3},-\sqrt{3})$ &  $\mathcal{Q}_i = (0, 1, 1)$ & anti-LRS Kasner solution\\
$\mathrm{S}_{\mathcal{T}_i} = \{\mathrm{S}_i\} \times \{\mathrm{T}_i\}$ &  
$ \mathrm{S}_i = (0,-\sqrt{3},+\sqrt{3})$ &  $\mathcal{T}_i = (1, 0, 0)$ & anti-LRS Kasner solution\\
$\mathrm{S}_{\mathcal{Q}_i} = \{\mathrm{S}_i\} \times \{\mathrm{Q}_i\}$ &  
$ \mathrm{S}_i = (0,-\sqrt{3},+\sqrt{3})$ &  $\mathcal{Q}_i = (0, 1, 1)$ & anti-LRS Kasner solution\\\hline
\end{tabular}
\end{center}

The boundary of ${}^{\mathrm{a}}\mathrm{LRS}_i$ forms a heteroclinic cycle,
\begin{equation}\label{antiheterocycle}
\begin{CD}
\mathrm{S}_{\mathcal{Q}_i} @<<< \mathrm{P}_{\mathcal{Q}_i} \\
@VVV @AAA \\
\mathrm{S}_{\mathcal{T}_i} @>>> \mathrm{P}_{\mathcal{T}_i}
\end{CD}
\end{equation}
and in the interior of the space there is the fixed point $\mathrm{F}$, cf.~\eqref{heterocycle}.

The global dynamics of anti-LRS solutions is therefore reminiscent of the dynamics of 
LRS solutions. We merely state the results; the proofs are analogous to
the proofs of Section~\ref{LRSsolutions}:
Each anti-LRS solution isotropizes toward the future. 
The eigenvalues of the linearization of the dynamical system at the 
point $\mathrm{F}$ are the same as in the LRS case; therefore we distinguish
two kinds of isotropization: Monotonic isotropization and oscillatory
isotropization. 
Toward the singularity
we observe oscillatory behavior between the two anti-LRS Kasner states;
this regime is connected with energy condition violation.

\section{Past asymptotic states}
\label{past}

This section is devoted to investigating the dynamics of 
general diagonal Bianchi type I 
solutions toward the initial singularity.
Theorem~\ref{pastthm} states that 
(i) if the elastic matter satisfies $a \in[-1,1)$
(which we will assume from now on)
and (ii) if we exclude the isotropic FRW solution
represented by $\mathrm{F}$ from our considerations,
then the $\alpha$-limit set
of every orbit in the state space $\mathcal{X}$ 
is located on $\partial\mathcal{X}$.
Hence, in order to understand the structure of the $\alpha$-limit set, it
is necessary to study in detail the flow on the boundary.
As a preparatory step, we discuss the network of fixed points (which
includes the Kasner circles) that is present
on $\partial \mathcal{X}$.
In Subsection~\ref{invsubsets}, we analyze in a step-by-step manner the flow on the invariant
subsets of $\partial\mathcal{X}$; in Subsection~\ref{structuresonb} these results are combined to identify certain special
structures on $\partial\mathcal{X}$, namely heteroclinic cycles and heteroclinic sequences.
Finally, in Subsection~\ref{alphalim} we condense the collected results into
statements (such as Theorem~\ref{genthm}) and conjectures 
on the possible $\alpha$-limit sets on $\partial\mathcal{X}$.

Since $\mathcal{X} = \mathit{\Sigma}\times \mathcal{Y}$, the boundary $\partial\mathcal{X}$
consists of two components, 
\begin{equation}\label{boundarystatespaceagain}
\partial \mathcal{X} = \left(\partial \mathit{\Sigma} \times \overline{\mathcal{Y}} \right) 
\cup
\left(\mathit{\overline{\Sigma}} \times \partial \mathcal{Y} \right)\:,
\end{equation}
the intersection of which is the set $\partial\mathit{\Sigma} \times \partial\mathcal{Y}$.

\begin{figure}[Ht]
\begin{center}
\psfrag{QL231}[bc][bc][0.8][60]{$\mathrm{QL}_{[2\bm{3}1]}$}
\psfrag{TL231}[bc][bc][0.8][60]{$\mathrm{TL}_{[2\bm{3}1]}$}
\psfrag{QL213}[bc][bc][0.8][0]{$\mathrm{QL}_{[2\bm{1}3]}$}
\psfrag{TL213}[bc][bc][0.8][0]{$\mathrm{TL}_{[2\bm{1}3]}$}
\psfrag{QL123}[bc][bc][0.8][-60]{$\mathrm{QL}_{[1\bm{2}3]}$}
\psfrag{TL123}[bc][bc][0.8][-60]{$\mathrm{TL}_{[1\bm{2}3]}$}
\psfrag{QL132}[tc][tc][0.8][60]{$\mathrm{QL}_{[1\bm{3}2]}$}
\psfrag{TL132}[tc][tc][0.8][60]{$\mathrm{TL}_{[1\bm{3}2]}$}
\psfrag{QL312}[tc][tc][0.8][0]{$\mathrm{QL}_{[3\bm{1}2]}$}
\psfrag{TL312}[tc][tc][0.8][0]{$\mathrm{TL}_{[3\bm{1}2]}$}
\psfrag{QL321}[tc][tc][0.8][-60]{$\mathrm{QL}_{[3\bm{2}1]}$}
\psfrag{TL321}[tc][tc][0.8][-60]{$\mathrm{TL}_{[3\bm{2}1]}$}
\psfrag{KCt1}[tc][tc][1.2][0]{$\mathrm{KC}_{\mathcal{T}_1}$}
\psfrag{KCq2}[tc][tc][1.2][0]{$\mathrm{KC}_{\mathcal{Q}_2}$}
\psfrag{KCt3}[tc][tc][1.2][0]{$\mathrm{KC}_{\mathcal{T}_3}$}
\psfrag{KCq1}[tc][tc][1.2][0]{$\mathrm{KC}_{\mathcal{Q}_1}$}
\psfrag{KCt2}[tc][tc][1.2][0]{$\mathrm{KC}_{\mathcal{T}_2}$}
\psfrag{KCq3}[tc][tc][1.2][0]{$\mathrm{KC}_{\mathcal{Q}_3}$}
\psfrag{sig1}[tc][tc][1][0]{$\mathrm{\Sigma}_1$}
\psfrag{sig2}[tc][tc][1][0]{$\mathrm{\Sigma}_2$}
\psfrag{sig3}[tc][tc][1][0]{$\mathrm{\Sigma}_3$}
\includegraphics[width=0.7\textwidth]{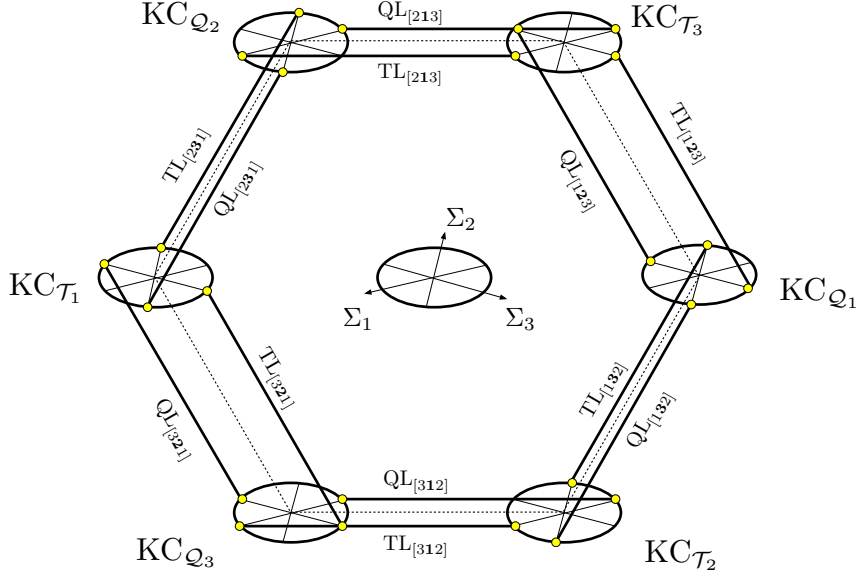}
\end{center}
\caption{A schematic depiction of the fixed points on the boundary $\partial\mathcal{X}$.}
\label{fixedpointspicture}
\end{figure}

By construction, the boundary sets $\partial \mathit{\Sigma} \times \overline{\mathcal{Y}}$
and $\mathit{\overline{\Sigma}} \times \partial \mathcal{Y}$ are invariant under
the flow of the (induced) dynamical system~\eqref{dynamical}.
There exists a number of equilibrium points of the dynamical system; these fixed points
are located not in the interior of the boundary components 
(i.e., neither on $\partial\mathit{\Sigma} \times \mathcal{Y}$
nor on $\mathit{\Sigma} \times \partial\mathcal{Y}$),
but on the shared boundary $\partial\mathit{\Sigma} \times \partial\mathcal{Y}$.
In fact, on $\partial\mathit{\Sigma} \times \partial\mathcal{Y}$ there exists
a connected network of one-parameter families of equilibrium points;
note in particular that there do not exist isolated fixed points.
\begin{itemize}
\item \textit{Kasner circles}: There exist six families of fixed points
that can be interpreted as Kasner circles (and are thus associated with
the Kasner solutions). They arise at
the six vertices $\{\mathcal{T}_1, \mathcal{T}_2,\mathcal{T}_3, \mathcal{Q}_1,\mathcal{Q}_2,\mathcal{Q}_3\}$ 
of $\partial\mathcal{Y}$, i.e.,
we have 
\begin{equation}\label{KCfam}
\mathrm{KC}_{\mathcal{T}_k} := \partial\mathit{\Sigma}\times \{\mathcal{T}_k\} \quad\: (k=1,2,3)\:,
\qquad\quad
\mathrm{KC}_{\mathcal{Q}_i}:= \partial \mathit{\Sigma}\times \{\mathcal{Q}_i\} \quad\:  (i=1,2,3)\:.
\end{equation}
\item \textit{Taub lines} and \textit{non-flat LRS lines}:
There exist
two lines of fixed points associated with each edge ($=$ sector) 
of $\partial\mathcal{Y}$. 
Consider the sector $[ijk] = 
\{(y_1,y_2,y_3)\;|\; y_i = 0,\, 0\leq y_j\leq1,\, y_k=1\}$; then 
the Taub line and the non-flat LRS line associated with $[ijk]$ are given by 
\begin{equation}\label{TQL}
\mathrm{TL}_{[i\bm{j}k]} := \{\mathrm{T}_j \} \times [ijk] \:,\qquad\quad
\mathrm{QL}_{[i\bm{j}k]} :=\{\mathrm{Q}_j \}\times [ijk]\:,
\end{equation}
respectively.
(For later purposes, we also introduce the 
associated open sets $\mathrm{TL}_{(i\bm{j}k)}$ and $\mathrm{QL}_{(i\bm{j}k)}$.
While $\mathrm{TL}_{[i\bm{j}k]} =  \{\mathrm{T}_j \} \times [ijk]$,
where $[ijk] = \{ 0 = y_i \leq y_j \leq y_k = 1\}$,
$\mathrm{TL}_{(i\bm{j}k)}$ is given by 
$\mathrm{TL}_{(i\bm{j}k)} = \{\mathrm{T}_j \} \times (ijk)$, where $(ijk) = \{ 0 = y_i < y_j < y_k =1\}$,
and analogously for $\mathrm{QL}_{(i\bm{j}k)}$).
\item Since $[ijk]$ connects $\mathcal{Q}_i$ with $\mathcal{T}_k$,
the intersection of these families of fixed points~\eqref{KCfam} and~\eqref{TQL} 
consists of 24 special points, which
are
$\{\mathrm{T}_j\} \times \{\mathcal{T}_k\}$, $\{\mathrm{Q}_j\} \times \{\mathcal{T}_k\}$,
$\{\mathrm{T}_j\} \times \{\mathcal{Q}_i\}$, $\{\mathrm{Q}_j\} \times \{\mathcal{Q}_i\}$,
where $i\neq j\neq k$.
\end{itemize}
This fixed point structure on the boundary of the state space $\partial\mathcal{X}$ 
is depicted in Figure~\ref{fixedpointspicture}.

For the subsequent analysis of the flow on the components
of $\partial\mathcal{X}$ we make the assumption $a>-1$.
The case $a=-1$ is degenerate in the sense that
the linearized dynamical system at certain fixed points
vanishes (i.e., all eigenvalues are zero).
Since this leads to some technical difficulties that we do not want to
discuss here, we henceforth assume $a\in (-1,1)$ (and $a b > 0$).
However, we note without proof 
that the key statements we derive in the following 
apply to the case $a=-1$ as well.

\subsection{Invariant subsets}
\label{invsubsets}


\subsubsection*{The boundary component $\bm{\partial\mathit{\Sigma}\times \overline{\mathcal{Y}}}$}

The set $\partial\mathit{\Sigma}\times \overline{\mathcal{Y}}$ is the subset
of $\overline{\mathcal{X}}$ characterized by $\Omega = 0$ (or, equivalently, $\Sigma^2 =1$).
The induced dynamical system on $\partial\mathit{\Sigma}\times \overline{\mathcal{Y}}$
reads
\begin{equation}\label{tor}
\Sigma_i^\prime = 0 \:,\qquad
y_i^\prime = -2 \,\epsilon_{ijk}\, y_i(1-y_i)\left[\Sigma_j-\Sigma_k\right] \quad(\text{no summation})\,,
\:\qquad (i=1,2,3) \:;
\end{equation}
in particular, $\Sigma_i =\mathrm{const}$ for all $i$.
Since $\Omega = 0$ and $\Sigma_i = \mathrm{const}$,
orbits on $\partial\mathit{\Sigma}\times \overline{\mathcal{Y}}$ represent
vacuum solutions (Kasner solutions); accordingly,
the set $\partial\mathit{\Sigma}\times \overline{\mathcal{Y}}$ can be
called the vacuum subset.

The space $\partial\mathit{\Sigma}\times \overline{\mathcal{Y}}$ can be
depicted as (set of points contained within or lying on) a 
torus whose cross section is 
the hexagon $\overline{\mathcal{Y}}$.
Since $\Sigma_i = \mathrm{const}$ for all $i$, 
each cross section $\{(\Sigma_1,\Sigma_2,\Sigma_3)\} \times \overline{\mathcal{Y}}$ 
is an invariant subspace.
Let $(\Sigma_1,\Sigma_2,\Sigma_3)$ be a element of sector $\langle ijk \rangle$ 
of $\partial\mathit{\Sigma}$, i.e., $\Sigma_i < \Sigma_j < \Sigma_k$.
Then $y_i^\prime \propto \epsilon_{ijk}$,
$y_j^\prime \propto (-\epsilon_{ijk})$, $y_k^\prime \propto \epsilon_{ijk}$,
hence the variables $y_l$ are increasing or decreasing for all $l$.
If $\epsilon_{ijk} =1$, the 
$\alpha$-limit of each orbit 
(in the interior of the space, i.e., in $\{(\Sigma_1,\Sigma_2,\Sigma_3)\} \times\mathcal{Y}$\,)
is the point $\{(\Sigma_1,\Sigma_2,\Sigma_3)\} \times \{\mathcal{T}_j\}$
and the $\omega$-limit is
the point  $\{(\Sigma_1,\Sigma_2,\Sigma_3)\} \times \{\mathcal{Q}_j\}$;
if $\epsilon_{ijk} = -1$, the roles of the points are interchanged.

When $(\Sigma_1,\Sigma_2,\Sigma_3)$ is one the Taub points $\mathrm{T}_j$,
then orbits
in the interior space $\{\mathrm{T}_j\} \times\mathcal{Y}$
emanate from a fixed point on the Taub line 
$\mathrm{TL}_{(k\bm{j}i)} = \{\mathrm{T}_j\} \times (kji)$,
where $\epsilon_{ijk} = 1$,
and end at a fixed point on the Taub line $\mathrm{TL}_{(i\bm{j}k)} =  \{\mathrm{T}_j\} \times (ijk)$.
The result is converse, when $(\Sigma_1,\Sigma_2,\Sigma_3)$ is one
of the non-flat LRS points $\mathrm{Q}_j$.
In that case, orbits
in the interior space $\{\mathrm{Q}_j\} \times\mathcal{Y}$
connect a fixed point on the non-flat LRS line $\mathrm{QL}_{(i\bm{j}k)} = \{\mathrm{Q}_j\} \times (ijk)$,
where $\epsilon_{ijk} = 1$,
with a fixed point on the Taub line $\mathrm{QL}_{(k\bm{j}i)} =  \{\mathrm{Q}_j\} \times (kji)$.

\begin{figure}[Ht]
\begin{center}
\psfrag{sig1}[tc][tc][1][0]{$\mathrm{\Sigma}_1$}
\psfrag{sig2}[tc][tc][1][0]{$\mathrm{\Sigma}_2$}
\psfrag{sig3}[tc][tc][1][0]{$\mathrm{\Sigma}_3$}
\psfrag{y1}[tc][tc][1][0]{$y_1$}
\psfrag{y2}[tc][tc][1][0]{$y_2$}
\psfrag{y3}[tc][tc][1][0]{$y_3$}
\psfrag{231}[lt][lt][0.8][0]{$\langle 231 \rangle$}
\psfrag{213}[tc][tc][0.8][0]{$\langle 213 \rangle$}
\psfrag{123}[rt][rt][0.8][0]{$\langle 123 \rangle$}
\psfrag{132}[rb][rb][0.8][0]{$\langle 132 \rangle$}
\psfrag{312}[bc][bc][0.8][0]{$\langle 312 \rangle$}
\psfrag{321}[lb][lb][0.8][0]{$\langle 321 \rangle$}
\psfrag{T1}[cc][cc][0.6][0]{$\mathrm{T}_1$}
\psfrag{T2}[cc][cc][0.6][0]{$\mathrm{T}_2$}
\psfrag{T3}[cc][cc][0.6][0]{$\mathrm{T}_3$}
\psfrag{Q1}[cc][cc][0.6][0]{$\mathrm{Q}_1$}
\psfrag{Q2}[cc][cc][0.6][0]{$\mathrm{Q}_2$}
\psfrag{Q3}[cc][cc][0.6][0]{$\mathrm{Q}_3$}
\psfrag{t1}[cc][cc][0.6][0]{$\mathcal{T}_1$}
\psfrag{t2}[cc][cc][0.6][0]{$\mathcal{T}_2$}
\psfrag{t3}[cc][cc][0.6][0]{$\mathcal{T}_3$}
\psfrag{q1}[cc][cc][0.6][0]{$\mathcal{Q}_1$}
\psfrag{q2}[cc][cc][0.6][0]{$\mathcal{Q}_2$}
\psfrag{q3}[cc][cc][0.6][0]{$\mathcal{Q}_3$}
\includegraphics[width=0.7\textwidth]{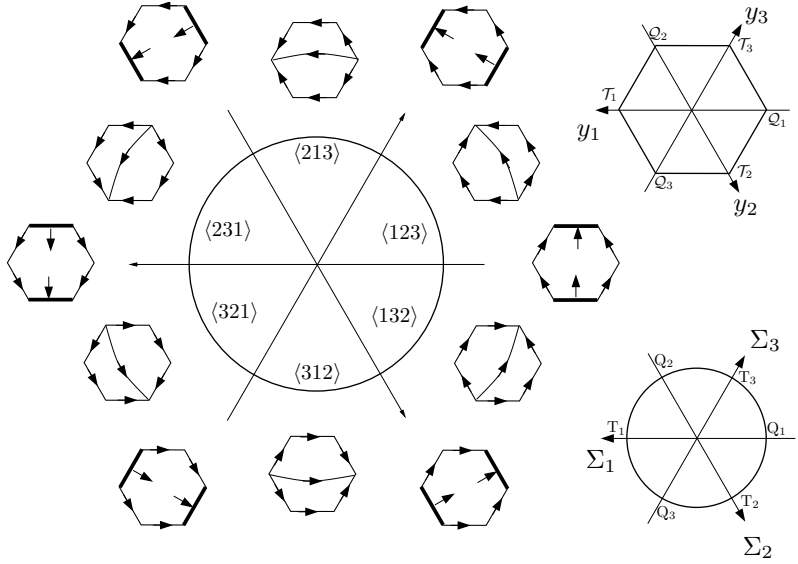}
\end{center}
\caption{Flow on the boundary component $\partial\mathit{\Sigma}\times \overline{\mathcal{Y}}$.
The cross sections of the ``torus'' $\partial\mathit{\Sigma}\times \overline{\mathcal{Y}}$
are invariant subspaces, since $(\Sigma_1,\Sigma_2,\Sigma_3) = \mathrm{const}$.
The flow on the section $\{(\Sigma_1,\Sigma_2,\Sigma_3)\} \times \overline{\mathcal{Y}}$ depends
on the position (sector) of $(\Sigma_1,\Sigma_2,\Sigma_3)$ on $\partial \mathit{\Sigma}$.
Note that all fixed points on $\partial\mathit{\Sigma}\times \overline{\mathcal{Y}}$ are 
hyperbolic or transversally hyperbolic.} 
\label{vacuum}
\end{figure}

The flow on the set $\partial\mathit{\Sigma}\times \overline{\mathcal{Y}}$
is depicted in Figure~\ref{vacuum}.

\subsubsection*{The boundary component $\bm{\overline{\mathit{\Sigma}}\times \partial\mathcal{Y}}$}

\begin{figure}[Ht]
\begin{center}
\psfrag{Q}[lc][lc][1.2][0]{$\overline{\mathit{\Sigma}}\times \mathcal{Q}_i$}
\psfrag{T}[lc][lc][1.2][0]{$\overline{\mathit{\Sigma}}\times \mathcal{T}_k$}
\psfrag{t}[rc][rc][1.2][0]{$\mathrm{TL}_{[i\bm{j}k]}$}
\psfrag{q}[lc][lc][1.2][0]{$\mathrm{QL}_{[i\bm{j}k]}$}
\psfrag{KC1}[lt][lt][1.2][0]{$\mathrm{KC}_{\mathcal{T}_k}$}
\psfrag{KC2}[lb][lb][1.2][0]{$\mathrm{KC}_{\mathcal{Q}_i}$}
\psfrag{y1}[cr][cr][0.7][0]{$\begin{pmatrix} y_i = 0 \\ y_j = 1\\ y_k = 1 \end{pmatrix}$}
\psfrag{y2}[cr][cr][0.7][0]{$\begin{pmatrix} y_i = 0 \\ y_j = 0\\ y_k = 1 \end{pmatrix}$}
\psfrag{y}[cr][cr][0.7][0]{$\begin{pmatrix} y_i = 0 \\ y_j \: \uparrow \\ y_k = 1 \end{pmatrix}$}
\includegraphics[width=0.45\textwidth]{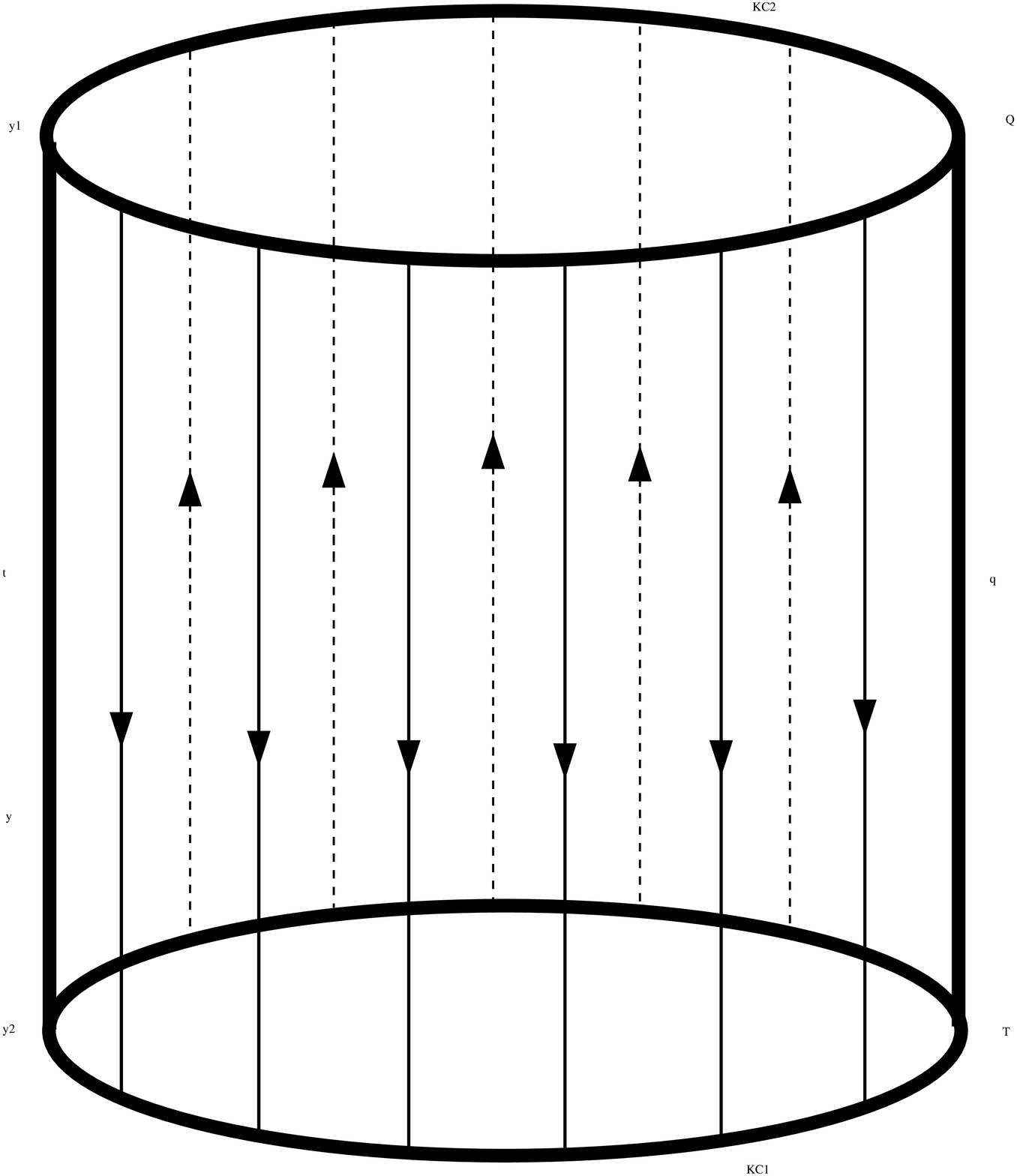}
\caption{The boundary component $\overline{\mathit{\Sigma}}\times \partial\mathcal{Y}$ 
consists of six cylinders $\mathrm{Cyl}_{[i\bm{j}k]}$. 
The top of $\mathrm{Cyl}_{[i\bm{j}k]} = \overline{\mathit{\Sigma}} \times [ijk]$
is the Kasner disc $\overline{\mathit{\Sigma}}\times \mathcal{Q}_i$, the base of $\mathrm{Cyl}_{[i\bm{j}k]}$
is the Kasner disc $\overline{\mathit{\Sigma}}\times \mathcal{T}_k$.
In this figure, $\mathrm{Cyl}_{[i\bm{j}k]}$ (with $\epsilon_{ijk} =1$) 
is depicted together with the flow of
the dynamical system on the lateral boundary.}
\label{cylinder}
\end{center}
\end{figure}

Since $\partial\mathcal{Y}$ consists of the six sectors $[ijk]$, see Figure~\ref{SigmaandY},
the set $\overline{\mathit{\Sigma}}\times \partial\mathcal{Y}$ can be viewed
as the union 
\begin{equation}\label{sixcyl}
\overline{\mathit{\Sigma}}\times \partial\mathcal{Y} = 
\bigcup_{ijk} \,\big( \overline{\mathit{\Sigma}}\times [ijk] \big)
=
\bigcup_{ijk} \,\mathrm{Cyl}_{[i\bm{j}k]}\:.
\end{equation}
When written out explicitly, we see that the set 
\begin{equation}
\mathrm{Cyl}_{[i\bm{j}k]}  = 
\overline{\mathit{\Sigma}} \times \big\{(y_1,y_2,y_3)\:|\: y_i = 0,\, 0\leq y_j\leq 1,\, y_k = 1\big\}
= \overline{\mathit{\Sigma}}\times [ijk]
\end{equation}
represents a cylinder, see Figure~\ref{cylinder}.
The six cylinders $\mathrm{Cyl}_{[i\bm{j}k]}$ are aligned along the hexagon $\partial\mathcal{Y}$, 
where each vertex corresponds to the top/base of a cylinder: $\overline{\mathit{\Sigma}}\times \mathcal{Q}_i$
($i=1,2,3$)
and $\overline{\mathit{\Sigma}}\times \mathcal{T}_k$ ($k=1,2,3$), respectively.

The flow on the space $\mathrm{Cyl}_{[i\bm{j}k]}$ is given by the induced dynamical system
\begin{subequations}\label{flowonCyl}
\begin{align}
\Sigma_i^\prime & = - 3\Omega \left[\frac{1}{2} (1-a) \Sigma_i - 2 \epsilon_{ijk} (1-y_j) \right]
& & y_i \equiv 0 \\
\Sigma_j^\prime & = - 3\Omega \left[\frac{1}{2} (1-a) \Sigma_j + 2 \epsilon_{ijk}\right]
& & y_j^\prime = -2 \epsilon_{ijk} y_j (1- y_j) (\Sigma_k - \Sigma_i) \\
\Sigma_k^\prime & = - 3\Omega \left[\frac{1}{2} (1-a) \Sigma_k - 2 \epsilon_{ijk} y_j \right]
& & y_k \equiv 1\:.
\end{align}
\end{subequations}

Since $(1-a) |\Sigma_j| < 4$ 
in the interior of $\mathrm{Cyl}_{[i\bm{j}k]}$
and in the interior of the top and the base of the cylinder
(i.e., on $\mathit{\Sigma}\times [ijk]$),
the derivative of $\Sigma_j$ has a sign, so that
$\Sigma_j$ is a monotone function.
The monotonicity principle thus implies that the $\alpha$- and
the $\omega$-limits of orbits must be located on the lateral
surface of $\mathrm{Cyl}_{[i\bm{j}k]}$ (which includes the
Kasner circles at the base/top).

The flow on the lateral surface of $\mathrm{Cyl}_{[i\bm{j}k]}$
is simple.
Since $(\Sigma_i,\Sigma_j,\Sigma_k) = \mathrm{const}$,
the equation $y_j^\prime = -2 \epsilon_{ijk} y_j (1-y_j) (\Sigma_k-\Sigma_i)$
contains the entire dynamical information.
Suppose that $\epsilon_{ijk} = 1$. Then, for $\mathrm{Cyl}_{[i\bm{j}k]}$
(and identically for $\mathrm{Cyl}_{[k\bm{j}i]}$) we find
$y_j^\prime  \gtrless 0$ when $\Sigma_i\gtrless \Sigma_k$.
Furthermore, $y_j^\prime = 0$ when $\Sigma_i = \Sigma_k$, which
is the case at the lines of fixed points $\mathrm{TL}_{[i\bm{j}k]}$ and
$\mathrm{QL}_{[i\bm{j}k]}$.
Therefore, the lateral surface consists of two domains where
$y_j$ is increasing/decreasing and which are separated by
the Taub line and the non-flat LRS line; see Figure~\ref{cylinder}.

This analysis leaves the equilibrium points, i.e.,
the sets $\mathrm{TL}_{[i\bm{j}k]}$,
$\mathrm{QL}_{[i\bm{j}k]}$, $\mathrm{KC}_{\mathcal{Q}_i}$, $\mathrm{KC}_{\mathcal{T}_k}$
as the only possible 
$\alpha$- and $\omega$-limit sets of orbits in $\mathrm{Cyl}_{[i\bm{j}k]}$.
It thus merely remains to identify those equilibrium points
that act as sources/sinks for
interior orbits.

To this end we make use of the auxiliary equations~\eqref{equationomega}; we find that 
\begin{equation}
\Omega^{-1} \Omega^\prime = 3(1-a) + 2\epsilon_{ijk} \left[ (1-y_j) (\Sigma_j-\Sigma_i) + y_j (\Sigma_j-\Sigma_k) \right]
\end{equation}
when evaluated at a fixed point on the lateral surface.
Consequently, for the equilibrium points we obtain
\begin{subequations}
\begin{alignat}{2}
& \mathrm{TL}_{[i\bm{j}k]}: &  \quad & 
\Omega^{-1} \Omega^\prime = \frac{3}{2} \Big[ (1 + \epsilon_{ijk}) (3-a) -(1-\epsilon_{ijk}) (1+a) \Big]\,, \\
& \mathrm{QL}_{[i\bm{j}k]}: & & 
\Omega^{-1} \Omega^\prime = \frac{3}{2} \Big[ (1 - \epsilon_{ijk}) (3-a) -(1+\epsilon_{ijk}) (1+a) \Big]\,, \\
\label{KCQi}
& \mathrm{KC}_{\mathcal{Q}_i}: & & 
\Omega^{-1} \Omega^\prime = 2 \Big[ \frac{3}{2} (1-a) + \epsilon_{ijk} (\Sigma_j - \Sigma_k) \Big]\,,
\\
\label{KCTk}
& \mathrm{KC}_{\mathcal{T}_k}: & &
\Omega^{-1} \Omega^\prime = 2 \Big[ \frac{3}{2} (1-a) + \epsilon_{ijk} (\Sigma_j - \Sigma_i) \Big]\,.
\end{alignat}
\end{subequations}
Suppose that $\epsilon_{ijk} = 1$ [$\epsilon_{ijk} = -1$].
Then each fixed point on $\mathrm{TL}_{(i\bm{j}k)}$ acts 
as a source [sink] for one interior orbit.
(Since the points on $\mathrm{TL}_{(i\bm{j}k)}$ and $\mathrm{QL}_{(i\bm{j}k)}$
are not transversally hyperbolic, this terminology is to be understood in
a broad sense. The precise statement reads:
Each fixed point on $\mathrm{TL}_{(i\bm{j}k)}$ acts as
the $\alpha$-limit [$\omega$-limit] for one interior orbit,
while there do not exist any other orbits that converge to this point
as $\tau\rightarrow\pm\infty$.)
Analogously,
the points on $\mathrm{QL}_{(i\bm{j}k)}$ act as sinks [sources].

The relations~\eqref{KCQi} and~\eqref{KCTk}, in conjunction with the properties of the flow on the
lateral boundary, lead to the following classification of the fixed points on the Kasner circles:
A fixed point on $\mathrm{KC}_{\mathcal{Q}_i}$ is a (transversally hyperbolic)
\begin{subequations}\label{KCsourcesink}
\begin{alignat}{3}
\label{KCsourcesinkA}
& \text{source/sink} \quad & & \Leftrightarrow \quad
\epsilon_{ijk} (\Sigma_j - \Sigma_k) \gtrless -\frac{3}{2} (1-a) & & \quad\text{and}\quad
\epsilon_{ijk} (\Sigma_i - \Sigma_k) \lessgtr 0\:. \\
\intertext{A fixed point on $\mathrm{KC}_{\mathcal{T}_k}$ is a (transversally hyperbolic)}
& \text{source/sink} \quad & & \Leftrightarrow \quad
\epsilon_{ijk} (\Sigma_i - \Sigma_j) \lessgtr \frac{3}{2} (1-a) & & \quad\text{and}\quad
\epsilon_{ijk} (\Sigma_i - \Sigma_k) \gtrless 0\:.
\end{alignat}
\end{subequations}
The remaining equilibrium points on the Kasner circles 
do not attract interior orbits as $\tau\rightarrow \pm\infty$.
(Except for four special points, these fixed points are 
transversally hyperbolic saddles.)

\begin{figure}[Ht]
\begin{center}
\psfrag{Q}[lc][lc][1.2][0]{$\overline{\mathit{\Sigma}}\times \mathcal{Q}_i$}
\psfrag{T}[lc][lc][1.2][0]{$\overline{\mathit{\Sigma}}\times \mathcal{T}_k$}
\psfrag{t}[rc][rc][1.2][0]{$\mathrm{TL}_{[i\bm{j}k]}$}
\psfrag{q}[lc][lc][1.2][0]{$\mathrm{QL}_{[i\bm{j}k]}$}
\psfrag{KC1}[lt][lt][1.2][0]{$\mathrm{KC}_{\mathcal{T}_k}$}
\psfrag{KC2}[lb][lb][1.2][0]{$\mathrm{KC}_{\mathcal{Q}_i}$}
\psfrag{y1}[cr][cr][0.7][0]{$\begin{pmatrix} y_i = 0 \\ y_j = 1\\ y_k = 1 \end{pmatrix}$}
\psfrag{y2}[cr][cr][0.7][0]{$\begin{pmatrix} y_i = 0 \\ y_j = 0\\ y_k = 1 \end{pmatrix}$}
\psfrag{sigj}[cc][cc][0.7][0]{$\Sigma_j$}
\psfrag{sigi}[cc][cc][0.7][0]{$\Sigma_i$}
\psfrag{sigk}[cc][cc][0.7][0]{$\Sigma_k$}
\psfrag{yj}[cc][cc][0.7][0]{$y_j$}
\includegraphics[width=0.8\textwidth]{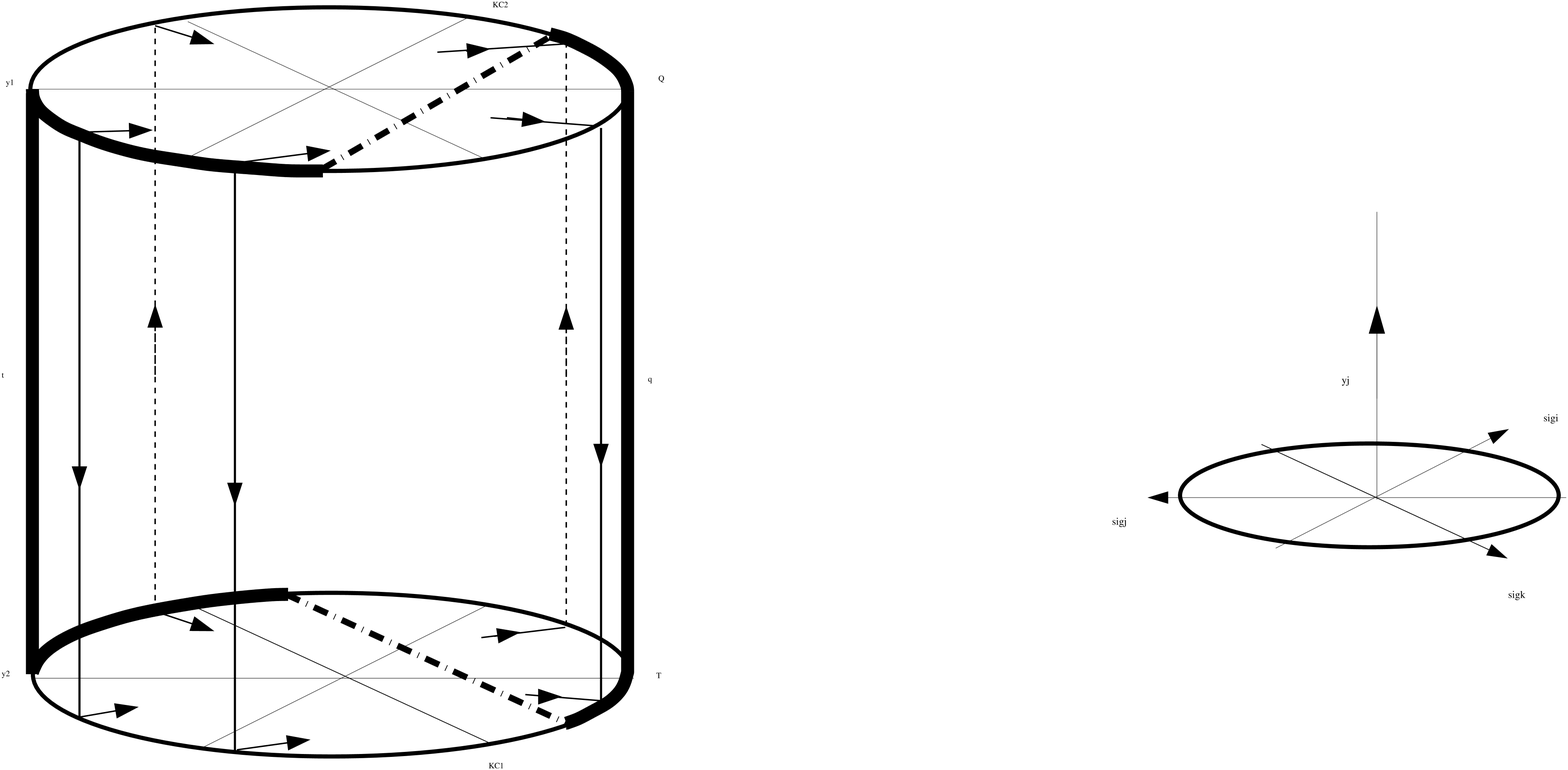}
\caption{A schematic depiction of the flow on
the cylinder $\mathrm{Cyl}_{[i\bm{j}k]} = \overline{\mathit{\Sigma}} \times [ijk]$ ($\epsilon_{ijk} = 1$).}
\label{cylinder2}
\end{center}
\end{figure}

The results of our analysis of the flow on the
set $\mathrm{Cyl}_{[i\bm{j}k]}$
are summarized in Figure~\ref{cylinder2}.
Each interior orbit (i.e., each orbit in the interior of $\mathrm{Cyl}_{[i\bm{j}k]}$)
is a heteroclinic orbit; it has a 
source as its $\alpha$-limit and a sink as its $\omega$-limit.

We conclude our study of the flow on the set $\mathrm{Cyl}_{[i\bm{j}k]}$
by noting that certain orbits can be given explicitly.
First, there is the solution
given by $\Sigma_i \equiv \Sigma_k$ and $y_j \equiv 1/2$,
which is $\mathrm{LRS}_j$; this solution
is a straight line that connects the fixed point
$\mathrm{T}_{(i\bm{j}k)} = \mathrm{T}_j \times \{(0,1/2,1)\}$ on $\mathrm{TL}_{[i\bm{j} k]}$
with the point $\mathrm{Q}_{(i\bm{j}k)} = \mathrm{Q}_j \times  \{(0,1/2,1)\}$ on $\mathrm{QL}_{[i\bm{j}k]}$,
see also~\eqref{heterocycle}.
Second, the orbits on the base/top of $\mathrm{Cyl}_{[i\bm{j}k]}$,
i.e., $\overline{\mathit{\Sigma}} \times \{\mathcal{Q}_i\}$
and $\overline{\mathit{\Sigma}} \times \{\mathcal{T}_k\}$,
are of a simple geometric form.

\begin{figure}[Ht]
\begin{center}
\psfrag{y1}[cc][cc][0.8][0]{$\Sigma_j$}
\psfrag{y2}[cc][cc][0.8][0]{$\Sigma_k$}
\psfrag{y3}[cc][cc][0.8][0]{$\Sigma_i$}
\psfrag{sigi0}[cc][cc][0.8][-30]{$\Sigma_i=0$}
\psfrag{sigk0}[cc][cc][0.8][30]{$\Sigma_k=0$}
\psfrag{sigqi}[cc][cc][1][0]{$\overline{\mathit{\Sigma}} \times \{\mathcal{Q}_i\}$}
\psfrag{sigtk}[cc][cc][1][0]{$\overline{\mathit{\Sigma}} \times \{\mathcal{T}_k\}$}
\includegraphics[width=0.8\textwidth]{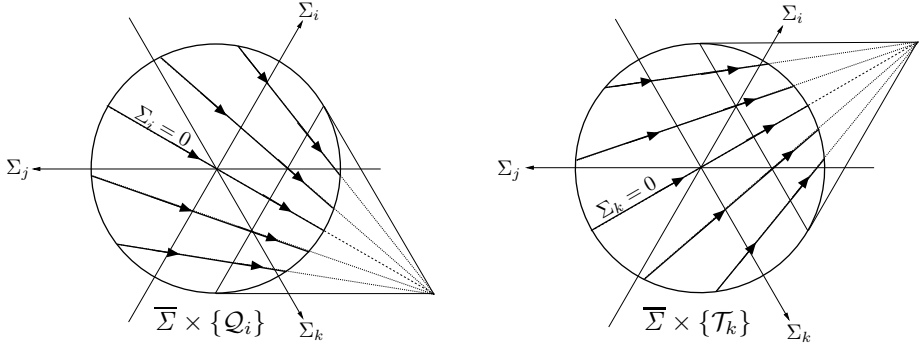}
\end{center}
\caption{Flow on the top (left) and base (right) of $\mathrm{Cyl}_{[i\bm{j}k]}$ ($\epsilon_{ijk}=1$).} \label{basetop}
\end{figure}

To see this, consider the top of $\mathrm{Cyl}_{[i\bm{j}k]}$, i.e., 
$\overline{\mathit{\Sigma}} \times \{\mathcal{Q}_i\}$. From the equations
\begin{equation}
\Sigma_i^\prime = - 3\Omega \left[\frac{1}{2} (1-a) \Sigma_i \right] \:,\qquad
\Sigma_{j/k}^\prime = - 3\Omega \left[\frac{1}{2} (1-a) \Sigma_{j/k} \pm 2 \epsilon_{ijk} \right]
\end{equation}
it is immediate that $\Sigma_i = 0$ is a solution of the system; clearly, this orbit 
is the intersection of $\mathrm{Cyl}_{[i\bm{j}k]}$ 
with the anti-LRS set ${}^{\mathrm{a}}\mathrm{LRS}_i$.
Furthermore, we find that
\begin{equation}\label{straightlines}
\frac{\frac{1}{2}(1-a)\Sigma_j + 2\epsilon_{ijk}}{\frac{1}{2}(1-a)\Sigma_k- 2\epsilon_{ijk}}= \mathrm{constant}
\end{equation}
under the flow of the system, whereby we obtain an explicit representation
of all orbits: Each orbit is a straight line. 
(Setting $\mathrm{const} = -1$ reproduces the anti-LRS orbit $\Sigma_i =0$.)
The one-parameter family of straight lines described by Equation~\eqref{straightlines}
possesses a common focal point, i.e., all lines intersect in the point
$(\Sigma_i, \Sigma_j,\Sigma_k) = \frac{4}{1-a} \epsilon_{ijk} (0, -1,1)$.
Two members of this family of straight lines are tangential to the Kasner circle
$\mathrm{KC}_{\mathcal{Q}_i}$. The two associated points of contact
lie on the straight line 
$\epsilon_{ijk} (\Sigma_j - \Sigma_k) = -\frac{3}{2} (1-a)$;
these fixed points are not transversally hyperbolic; see~\eqref{KCsourcesinkA}.
For a depiction of the flow on $\overline{\mathit{\Sigma}} \times \{\mathcal{Q}_i\}$
and the analogous flow on $\overline{\mathit{\Sigma}} \times \{\mathcal{T}_k\}$, see Figure~\ref{basetop}.

The three special orbits that exist on $\mathrm{Cyl}_{[i\bm{j}k]}$ are particularly relevant
for further purposes. Let us thus recapitulate:
There exists an anti-LRS orbit ${}^{\mathrm{a}}\mathrm{LRS}_i$ given
by $\Sigma_i = 0$ on the top of the cylinder,
an $\mathrm{LRS}_j$ orbit $\Sigma_i = \Sigma_k$ in the middle of $\mathrm{Cyl}_{[i\bm{j}k]}$, and an
anti-LRS orbit ${}^{\mathrm{a}}\mathrm{LRS}_k$ given by $\Sigma_k = 0$ on the bottom.

\subsection{Structures on the boundary}
\label{structuresonb}

A crucial property of the equilibrium points on $\partial\mathcal{X}$ is
that all these points are \textit{saddles}.
This is straightforward to prove:
Consider first a fixed point on $\mathrm{TL}_{[i\bm{j}k]} = \{\mathrm{T}_j\} \times [ijk]$.
As shown previously, 
such a point acts 
as a source [sink] within the subset 
$\mathrm{Cyl}_{[i\bm{j}k]} = \overline{\mathit{\Sigma}}\times [ijk]$ when $\epsilon_{ijk}= 1$ [$\epsilon_{ijk} = -1$].
However, within the (linearly independent) 
invariant subset $\{\mathrm{T}_j\} \times \overline{\mathcal{Y}}$
it is a (transversally hyperbolic) sink [source]. Consequently,
the point acts as a saddle. In particular, orbits in $\mathcal{X}$ cannot
converge to $\mathrm{TL}_{[i\bm{j}k]}$ as $\tau\rightarrow \pm \infty$.
The case $\mathrm{QL}_{[i\bm{j}k]}$ is analogous.

Consider now the Kasner circle $\mathrm{KC}_{\mathcal{Q}_i}$.
This circle is part of the boundary of two independent invariant subsets:
$\mathrm{Cyl}_{[i\bm{j}k]} = \overline{\mathit{\Sigma}} \times [ijk]$ 
and $\mathrm{Cyl}_{[i\bm{k}j]}=\overline{\mathit{\Sigma}} \times [ikj]$;
without loss of generality we assume $\epsilon_{ijk} = 1$.
From~\eqref{KCsourcesink} we conclude that a point on
\begin{subequations}
\begin{alignat}{2}
& \mathrm{KC}_{\mathcal{Q}_i} \subseteq \mathrm{Cyl}_{[i\bm{j}k]} 
\quad \text{is a source/sink} 
& \quad & \Leftrightarrow \quad 
(\Sigma_j - \Sigma_k) \gtrless -\frac{3}{2} (1-a)\,, \quad
(\Sigma_i - \Sigma_k) \lessgtr 0 \,, \\
& \mathrm{KC}_{\mathcal{Q}_i} \subseteq \mathrm{Cyl}_{[i\bm{k}j]} 
\quad \text{is a source/sink} 
& \quad & \Leftrightarrow \quad 
(\Sigma_j - \Sigma_k) \gtrless -\frac{3}{2} (1-a) \,, \quad
(\Sigma_j - \Sigma_i) \lessgtr 0\:.
\end{alignat}
\end{subequations}
Since these conditions are mutually exclusive, every equilibrium point
on $\mathrm{KC}_{\mathcal{Q}_i}$ acts as a saddle.
The analogous considerations apply to the case $\mathrm{KC}_{\mathcal{T}_k}$.

\subsubsection*{Heteroclinic cycles}

From the analysis of the flow on the invariant subsets of $\partial \mathcal{X}$
we infer the existence of a multitude of heteroclinic cycles.
On the one hand there are the three LRS cycles~\eqref{heterocycle}, which
arise as the intersection of $\partial\mathcal{X}$ with $\mathrm{LRS}_i$, $i=1,2,3$; see Section~\ref{LRSsolutions}.
On the other hand there exist the three anti-LRS cycles~\eqref{antiheterocycle},
which are the intersection of $\partial\mathcal{X}$ with ${}^{\mathrm{a}}\mathrm{LRS}_i$, $i=1,2,3$,
so that $\Sigma_i = 0$; see Section~\ref{antiLRSsolutions}.
There are, however, infinitely many heteroclinic cycles associated with the condition $\Sigma_i = 0$;
this will be discussed next.

On the boundary component $\partial\mathit{\Sigma} \times \overline{\mathcal{Y}}$
consider, without loss of generality, the subset $\Sigma_1 = 0$.
It consists of the two disconnected sets
$\{\mathrm{P}_1\} \times \overline{\mathcal{Y}}$ and $\{\mathrm{S}_1\} \times \overline{\mathcal{Y}}$,
where the point $\mathrm{P}_1 = (0, \sqrt{3}, -\sqrt{3})$ 
is the anti-LRS point of sector $\langle 312 \rangle$ 
and $\mathrm{S}_1 = (0, -\sqrt{3},\sqrt{3})$ the anti-LRS point of sector $\langle 213 \rangle$ 
of $\partial \mathit{\Sigma}$.
The $\alpha$-limit of each orbit in $\{\mathrm{P}_1\} \times \mathcal{Y}$ is the fixed point
$\{\mathrm{P}_1\} \times \{ \mathcal{T}_1\}$, the $\omega$-limit is 
$\{\mathrm{P}_1\} \times \{ \mathcal{Q}_1\}$, see Figure~\ref{vacuum}.
Note that these fixed points are not only connected with each other through
orbits in $\{\mathrm{P}_1\} \times \mathcal{Y}$, but also via the sequences of 
boundary orbits 
\begin{equation}\label{detours}
\begin{split}
& \{\mathrm{P}_1\} \times \{ \mathcal{T}_1\} \longrightarrow
\{\mathrm{P}_1\} \times \{ \mathcal{Q}_2\}  \longrightarrow
\{\mathrm{P}_1\} \times \{ \mathcal{T}_3\}  \longrightarrow
\{\mathrm{P}_1\} \times \{ \mathcal{Q}_1\} \\
\text{and} \quad & \{\mathrm{P}_1\} \times \{ \mathcal{T}_1\} \longrightarrow
\{\mathrm{P}_1\} \times \{ \mathcal{Q}_3\}  \longrightarrow
\{\mathrm{P}_1\} \times \{ \mathcal{T}_2\}  \longrightarrow
\{\mathrm{P}_1\} \times \{ \mathcal{Q}_1\}\:.
\end{split}
\end{equation}
For the subset $\{\mathrm{S}_1\} \times \mathcal{Y}$ the roles of the two fixed points
are reversed: $\{\mathrm{S}_1\} \times \{ \mathcal{Q}_1\}$ acts as $\alpha$-limit,
$\{\mathrm{S}_1\} \times \{ \mathcal{T}_1\}$ as $\omega$-limit.

The fixed point $\{\mathrm{P}_1\} \times \{ \mathcal{Q}_1\}$
is a saddle. While it acts as the $\omega$-limit for the one-parameter family of orbits 
on $\{\mathrm{P}_1\} \times \mathcal{Y}$, 
there exists exactly one orbit emanating from it: 
The ${}^{\mathrm{a}}\mathrm{LRS}_1$ orbit $\Sigma_1 = 0$ on $\mathit{\Sigma} \times \{ \mathcal{Q}_1\}$
connects $\{\mathrm{P}_1\} \times \{ \mathcal{Q}_1\}$ with 
$\{\mathrm{S}_1\} \times \{ \mathcal{Q}_1\}$.
Analogously, the ${}^{\mathrm{a}}\mathrm{LRS}_1$ orbit $\Sigma_1 = 0$ on 
$\mathit{\Sigma} \times \{ \mathcal{T}_1\}$
is the unique orbit connecting
$\{\mathrm{S}_1\} \times \{ \mathcal{T}_1\}$
with $\{\mathrm{P}_1\} \times \{ \mathcal{T}_1\}$.

\begin{figure}[Ht]
\begin{center}
\psfrag{p1+t1}[cc][cc][0.7][0]{$\{\mathrm{P}_1\} \times \{ \mathcal{T}_1\}$}
\psfrag{p1-t1}[cc][cc][0.7][0]{$\{\mathrm{S}_1\} \times \{ \mathcal{T}_1\}$}
\psfrag{p1+q2}[cc][cc][0.7][0]{$\{\mathrm{P}_1\} \times \{ \mathcal{Q}_2\}$}
\psfrag{p1-q2}[cc][cc][0.7][0]{$\{\mathrm{S}_1\} \times \{ \mathcal{Q}_2\}$}
\psfrag{p1+t3}[cc][cc][0.7][0]{$\{\mathrm{P}_1\} \times \{ \mathcal{T}_3\}$}
\psfrag{p1-t3}[cc][cc][0.7][0]{$\{\mathrm{S}_1\} \times \{ \mathcal{T}_3\}$}
\psfrag{p1+q1}[cc][cc][0.7][0]{$\{\mathrm{P}_1\} \times \{ \mathcal{Q}_1\}$}
\psfrag{p1-q1}[cc][cc][0.7][0]{$\{\mathrm{S}_1\} \times \{ \mathcal{Q}_1\}$}
\psfrag{p1+t2}[cc][cc][0.7][0]{$\{\mathrm{P}_1\} \times \{ \mathcal{T}_2\}$}
\psfrag{p1-t2}[cc][cc][0.7][0]{$\{\mathrm{S}_1\} \times \{ \mathcal{T}_2\}$}
\psfrag{p1+q3}[cc][cc][0.7][0]{$\{\mathrm{P}_1\} \times \{ \mathcal{Q}_3\}$}
\psfrag{p1-q3}[cc][cc][0.7][0]{$\{\mathrm{S}_1\} \times \{ \mathcal{Q}_3\}$}
\includegraphics[width=0.6\textwidth]{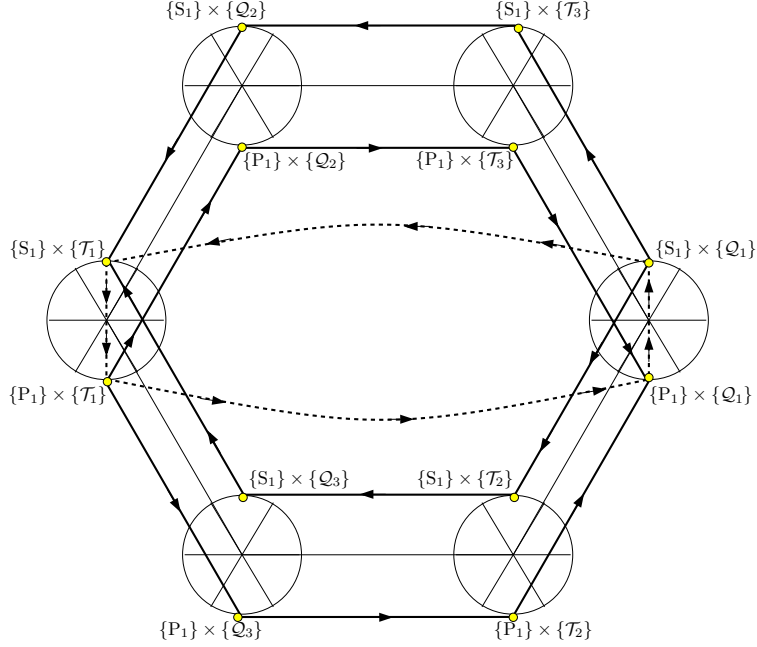}
\end{center}
\caption{A depiction of the heteroclinic cycles~\eqref{hetero2}; the dashed lines form a 
heteroclinic cycle of the family~(\ref{hetero}).}
\label{heteroclinicpicture}
\end{figure}

Therefore, we have found a two-parameter family of heteroclinic cycles
involving the fixed points $\{\mathrm{S}_1/\mathrm{P}_1\} \times \{ \mathcal{T}_1/\mathcal{Q}_1\}$,
which can be written down schematically as follows:
\begin{equation}\label{hetero}
\begin{CD}
\{\mathrm{S}_1\} \times \{ \mathcal{Q}_1\}    
@<\text{unique}<<   \{\mathrm{P}_1\} \times \{ \mathcal{Q}_1\} \\
@V\text{infinitely many}VV @AA\text{infinitely many}A \\
\{\mathrm{S}_1\} \times \{ \mathcal{T}_1\} 
@>>\text{unique}> \{\mathrm{P}_1\} \times \{ \mathcal{T}_1\}  
\end{CD}
\end{equation}
Evidently, this family of heteroclinic cycles represent a generalization of 
the anti-LRS cycle~\eqref{antiheterocycle}.
In fact, as noted in~\eqref{detours}, 
the flow on $\{\mathrm{S}_1/\mathrm{P}_1\} \times \overline{\mathcal{Y}}$
permits alternative completions of the heteroclinic cycles:
Possible ``detours'' include additional fixed points, see Figure \ref{heteroclinicpicture}.
\begin{equation}\label{hetero2}
\begin{CD}
\{\mathrm{S}_1\} \times \{\mathcal{T}_3/\mathcal{T}_2\} 
@<\text{detour}<<
\{\mathrm{S}_1\} \times \{\mathcal{Q}_1\}    
@<\text{unique}<<   
\{\mathrm{P}_1\} \times \{\mathcal{Q}_1\} 
@<\text{detour}<<
\{\mathrm{P}_1\} \times \{\mathcal{T}_3/\mathcal{T}_2\} 
\\
@V\text{detour}VV 
@V\text{infinitely}V\text{many}V 
@A\text{infinitely}A\text{many}A 
@AA\text{detour}A
\\
\{\mathrm{S}_1\} \times \{\mathcal{Q}_2/\mathcal{Q}_3\} 
@>>\text{detour}>
\{\mathrm{S}_1\} \times \{\mathcal{T}_1\} 
@>>\text{unique}> 
\{\mathrm{P}_1\} \times \{\mathcal{T}_1\} 
@>>\text{detour}>
\{\mathrm{P}_1\} \times \{\mathcal{Q}_2/\mathcal{Q}_3\} 
\end{CD}
\end{equation}

\subsubsection*{Sequences of heteroclinic orbits}

As we have seen in Subsection~\ref{invsubsets},
the flow on the subsets $\overline{\mathit{\Sigma}} \times \{ \mathcal{Q}_1\}$
and $\overline{\mathit{\Sigma}} \times \{ \mathcal{T}_1\}$
is represented by a family of straight lines that possess a common focal point;
see Figure~\ref{basetop}.
The anti-LRS line $\Sigma_1 = 0$ is the central line. 
Orbits that are close to $\Sigma_1 = 0$ (i.e., orbits where $\Sigma_1$ is small)
connect the same sectors as the central line, namely sector $\langle 312 \rangle$ with sector $\langle 213 \rangle$.
Let $\mathrm{P}_{\langle 312 \rangle}$ denote any point on sector $\langle 312 \rangle$ of $\partial\mathit{\Sigma}$
(the anti-LRS point $\mathrm{P}_1$ being an example) and likewise $\mathrm{P}_{\langle 213 \rangle}$ 
any point on sector $\langle 312 \rangle$
(such as $\mathrm{S}_1$).
Symbolically we then write for any orbit on $\overline{\mathit{\Sigma}} \times \{ \mathcal{Q}_1\}$
that is close to $\Sigma_1 = 0$
\begin{equation}
\mathrm{P}_{\langle 312 \rangle} \times \{ \mathcal{Q}_1\}  \longrightarrow \mathrm{P}_{\langle 213 \rangle} \times \{ \mathcal{Q}_1\}\:,
\end{equation}
and likewise for any orbit on  $\overline{\mathit{\Sigma}} \times \{ \mathcal{T}_1\}$.
For these
orbits it is immediate that $|\Sigma_1|$ is monotonically decreasing
in $\tau$. Accordingly, \textit{when we invert the direction of time},
$|\Sigma_1|$ is monotonically increasing along each orbit, so that
all orbits diverge from the central line $\Sigma_1 = 0$.

\begin{figure}[Ht]
\begin{center}
\psfrag{KCt1}[tc][tc][1.2][0]{$\mathrm{KC}_{\mathcal{T}_1}$}
\psfrag{KCq1}[tc][tc][1.2][0]{$\mathrm{KC}_{\mathcal{Q}_1}$}
\includegraphics[width=0.6\textwidth]{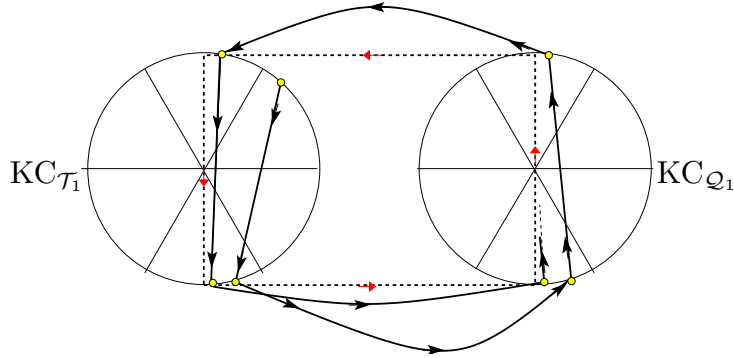}
\end{center}
\caption{Example of heteroclinic sequence~\eqref{heteroseq} diverging (as $\tau\to-\infty$) 
from a heteroclinic cycle of the family~\eqref{hetero} (represented by a dashed line).}
\label{heteroclinicsequencepicture}
\end{figure}

These orbits on $\overline{\mathit{\Sigma}} \times \{ \mathcal{Q}_1\}$
and $\overline{\mathit{\Sigma}} \times \{ \mathcal{T}_1\}$
can be used to construct sequences of heteroclinic orbits.
Such a sequence of heteroclinic orbits can be written schematically as follows:
\begin{equation}\label{heteroseq}
\begin{CD}
\{\mathrm{S}_{\langle 213 \rangle}\} \times \{\mathcal{T}_{3/2}\} 
@<\text{detour}<<
\{\mathrm{S}_{\langle 213 \rangle}\} \times \{\mathcal{Q}_1\}    
@<\text{unique}<<   
\{\mathrm{P}_{\langle 312 \rangle}\} \times \{\mathcal{Q}_1\} 
@<\text{detour}<<
\{\mathrm{P}_{\langle 312 \rangle}\} \times \{\mathcal{T}_{3/2}\} 
\\
@V\text{detour}VV 
@V\text{infinitely}V\text{many}V 
@A\text{infinitely}A\text{many}A 
@AA\text{detour}A
\\
\{\mathrm{S}_{\langle 213 \rangle}\} \times \{\mathcal{Q}_{2/3}\} 
@>>\text{detour}>
\{\mathrm{S}_{\langle 213 \rangle}\} \times \{\mathcal{T}_1\} 
@>>\text{unique}> 
\{\mathrm{P}_{\langle 312 \rangle}\} \times \{\mathcal{T}_1\} 
@>>\text{detour}>
\{\mathrm{P}_{\langle 312 \rangle}\} \times \{\mathcal{Q}_{2/3}\} 
\end{CD}
\end{equation}
Naturally, in the course of the sequence, the points $\mathrm{S}_{\langle 213 \rangle}$ and 
$\mathrm{P}_{\langle 312 \rangle}$ change; in particular, $|\Sigma_1|$ increases (with $-\tau$). 
An example of such behavior is depicted in Figure \ref{heteroclinicsequencepicture}. 

Since $|\Sigma_1|$ grows, any given sequence of the type~\eqref{heteroseq} will eventually leave
the sectors $\langle 213 \rangle$ and $\langle 312 \rangle$. 
There exist uncountably many possible continuations;
however, while~\eqref{heteroseq} is associated with a unique sequence
of Kasner states,
the possible continuations of a given sequence are not: They 
give rise to different sequences of Kasner states. 
Since this makes the analysis difficult, we refrain from
going into details.

\subsection{$\pmb{\alpha}$-limits}
\label{alphalim}

It is obvious that 
the complexity of the structures that are present on the boundary
of the state space impedes a rigorous mathematical derivation
of the past attractor of orbits in $\mathcal{X}$.
We confine ourselves to presenting some basic results.

Theorem~\ref{pastthm} implies that the $\alpha$-limit of every orbit in $\mathcal{X}$
must be a subset of $\partial\mathcal{X}$. Since all fixed points on $\partial\mathcal{X}$
act as saddles, 
there do not exist orbits in $\mathcal{X}$ whose $\alpha$-limit
set consists of merely a point.
In particular, there do not exist solutions that converge to a
Kasner solution as $t\rightarrow 0$.

The past asymptotics of LRS solutions and anti-LRS solutions 
has been analyzed in Sections~\ref{LRSsolutions} and~\ref{antiLRSsolutions}:
The asymptotic behavior of solutions is characterized by
oscillations between two different Kasner solutions.
For solutions that are neither LRS nor anti-LRS we have the
following theorem:

\begin{Theorem}\label{genthm}
The $\alpha$-limit set of a (non-LRS, non-${}^{\mathit{a}}\!\mathit{LRS}$) 
solution in $\mathcal{X}$ comprises a large (probably infinite) set of Kasner points.
\end{Theorem}

\textit{Interpretation of the theorem}.
The theorem states that, as $\tau\rightarrow -\infty$, the solution
undergoes an infinite sequence of phases (``epochs''), 
in each of which the behavior of the solution is approximately described by
a certain Kasner solution. We conjecture that the number of Kasner states 
the (generic) solution passes through in this way is infinite.

\begin{proof}
Consider an orbit $\gamma$ in $\mathcal{X}$ which is neither LRS nor ${}^{\mathrm{a}}\mathrm{LRS}$.
Suppose that a point $\mathcal{P}\in\partial\mathcal{X}$ is an element of $\alpha(\gamma)$.
Since $\mathcal{P}\in\alpha(\gamma)$, the orbit through $\mathcal{P}$ and
its $\alpha$-limit point $\mathcal{P}_-$ and its $\omega$-limit point
$\mathcal{P}_+$ must also lie in $\alpha(\gamma)$. (As a matter of course, 
since all orbits on $\partial\mathcal{X}$ are heteroclinic orbits, $\mathcal{P}_\pm$ are
fixed points.)
Since $\mathcal{P}_+\in\alpha(\gamma)$, there exists an orbit emanating from $\mathcal{P}_+$
that is contained in $\alpha(\gamma)$ as well; likewise, there
exists an orbit converging to $\mathcal{P}_-$ that lies in $\alpha(\gamma)$.
Continuing in this manner we can construct a sequence of fixed points 
and thus a sequence of associated Kasner states $\mathcal{K}_n$, $n\in\mathbb{Z}$,
that is contained in $\alpha(\gamma)$. 
Taking account of the previous analysis
of the flow on $\partial\mathcal{X}$ it is not difficult
to convince oneself that the sequence $\mathcal{K}_n$ does not exhibit
any simple recurrence (and probably no recurrence at all in the
generic case); this is, however, provided that
$\mathcal{P}\not\in \mathrm{LRS}_i$ and $\mathcal{P}\not\in {}^{\mathrm{a}}\mathrm{LRS}_i$
($i=1,2,3$). In that special case, the sequence $\mathcal{K}_n$ is
the alternating sequence of Kasner states described by~\eqref{heterocycle}
or~\eqref{antiheterocycle}, respectively.
In order to prove the theorem it thus remains to show that
the intersection of $\alpha(\gamma)$ with $\mathrm{LRS}_i$ 
and ${}^{\mathrm{a}}\mathrm{LRS}_i$ is empty, or, equivalently,
that $\gamma$ cannot converge to any of the heteroclinic cycles~\eqref{heterocycle} 
or~\eqref{antiheterocycle} as $\tau\rightarrow \infty$.
To see that we merely note that these heteroclinic cycles are not stable.
Consider, for instance, the LRS orbit in $\mathrm{Cyl}_{[i\bm{j}k]}$.
A small perturbation of that orbit results in different $\alpha$- and
$\omega$-limit (that are located on the Kasner circles, generically),
and thus leads to a quickly increasing deviation from LRS.
Analogously, consider the anti-LRS orbit $\Sigma_i = 0$ 
on $\overline{\mathit{\Sigma}}\times \{\mathcal{Q}_i\}$
or $\overline{\mathit{\Sigma}}\times \{\mathcal{T}_i\}$.
Since $|\Sigma_i|$ grows along the other orbits on 
$\overline{\mathit{\Sigma}}\times \{\mathcal{Q}_i/\mathcal{T}_i\}$,
perturbations of the anti-LRS orbit increase with increasing 
$(-\tau)$.
This establishes the claimed
instability of the LRS and ${}^{\mathrm{a}}\mathrm{LRS}$ cycles.
\end{proof}

\begin{figure}[Ht]
\begin{center}
\psfrag{y1}[cc][cc][1][0]{$y_1$}
\psfrag{y2}[cc][cc][1][0]{$y_2$}
\psfrag{y3}[cc][cc][1][0]{$y_3$}
\psfrag{231}[lt][lt][1][0]{$[231]$}
\psfrag{a231}[bc][bc][1][60]{$(1,0,\rightarrow)$}
\psfrag{213}[tc][tc][1][0]{$[213]$}
\psfrag{a213}[bc][bc][1][0]{$(\leftarrow,0,1)$}
\psfrag{123}[rt][rt][1][0]{$[123]$}
\psfrag{a123}[bc][bc][1][-60]{$(0,\rightarrow,1)$}
\psfrag{132}[rb][rb][1][0]{$[132]$}
\psfrag{a132}[tc][tc][1][60]{$(0,1,\rightarrow)$}
\psfrag{312}[bc][bc][1][0]{$[312]$}
\psfrag{a312}[tc][tc][1][0]{$(\leftarrow,1,0)$}
\psfrag{321}[lb][lb][1][0]{$[321]$}
\psfrag{a321}[tc][tc][1][-60]{$(1,\rightarrow,0)$}
\psfrag{T1}[cc][cc][1][0]{$\mathcal{T}_1$}
\psfrag{T2}[cc][cc][1][0]{$\mathcal{T}_2$}
\psfrag{T3}[cc][cc][1][0]{$\mathcal{T}_3$}
\psfrag{Q1}[cc][cc][1][0]{$\mathcal{Q}_1$}
\psfrag{Q2}[cc][cc][1][0]{$\mathcal{Q}_2$}
\psfrag{Q3}[cc][cc][1][0]{$\mathcal{Q}_3$}
\includegraphics[width=0.45\textwidth]{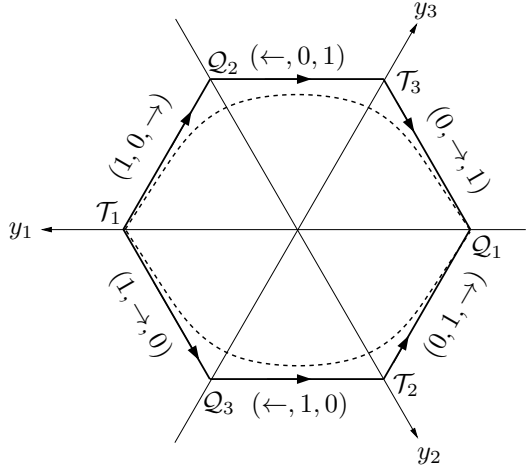}
\caption{An orbit converging to a heteroclinic sequence of the type~\eqref{heteroseq}
is increasingly forced to the boundary of the hexagon by~\eqref{y2y3}.
The dashed lines represent the curves $y_2/(1-y_2) = \mathrm{const}\: y_3/(1-y_3)$,
where the constant is either small (upper dashed line) or large (lower dashed line).}
\label{pathsonhexagon}
\end{center}
\end{figure}

Having shown in the proof of the theorem that there do not exist
orbits that converge to the LRS/${}^{\mathrm{a}}\mathrm{LRS}$ heteroclinic cycles
except for the LRS/${}^{\mathrm{a}}\mathrm{LRS}$ solutions themselves,
one follow-up question suggests itself:
Do there exist orbits that converge to a sequence of heteroclinic
orbits of the type~\eqref{heteroseq} described in the previous subsection?
To analyze this question, consider the heteroclinic sequence~\eqref{heteroseq}
and suppose that
there exists an orbit in $\mathcal{X}$ that
shadows this sequences in the asymptotic regime $\bar{\tau} = (-\tau) \rightarrow \infty$.
Along the orbit we thus have
\begin{equation}\label{tent}
\frac{d}{d\bar{\tau}}\left[ \log \frac{y_2/(1-y_2)}{y_3/(1-y_3)} \right] = -6 \Sigma_1 
\end{equation}
and accordingly
\begin{equation}\label{y2y3}
\frac{y_2}{1-y_2} \propto \frac{y_3}{1-y_3} \exp\left[ -6 \int \Sigma_1 d\bar{\tau} \right]\:.
\end{equation}
Since $\Sigma_1$ has a sign along the sequence of heteroclinic orbits,
$\int \Sigma_1 d\bar{\tau}$ is monotone in $\bar{\tau}$; in fact, since
the orbit stays a large amount of time close to the fixed points,
the (absolute value of) the integral is very large.
Consequently, \eqref{y2y3} implies that $y_2$ and $y_3$ 
cannot remain of the same order as $\bar{\tau}$ increases.
Therefore, in the asymptotic regime $\bar{\tau}\rightarrow \infty$,
the orbit is forced to the boundary of $\mathcal{Y}$,
see Figure~\ref{pathsonhexagon}. In the terminology
of the previous subsection we can say that
the orbit is forced on the detour paths of~\eqref{heteroseq}.

A detailed analysis (which we omit here) of the fixed points
$\{\mathrm{P}_{\langle 312 \rangle}\} \times \{\mathcal{Q}_{2/3}\}$
and $\{\mathrm{S}_{\langle 213 \rangle}\} \times \{\mathcal{T}_{3/2}\}$
reveals that additional branchings can occur at these points.
If $a \geq 1- 2/\sqrt{3}$, then 
$\{\mathrm{P}_{\langle 312 \rangle}\} \times \{\mathcal{Q}_{2/3}\}$
acts as the $\omega$-limit not only for one
orbit (the orbit 
$\{\mathrm{P}_{\langle 312 \rangle}\} \times \{\mathcal{T}_1\} 
\rightarrow 
\{\mathrm{P}_{\langle 312 \rangle}\} \times \{\mathcal{Q}_{2/3}\}$),
but for a one-parameter family of orbits. 
(In fact, $\{\mathrm{P}_{\langle 312 \rangle}\} \times \{\mathcal{Q}_{2/3}\}$ is a sink
in $\mathrm{Cyl}_{[2\bm{3}1]}$ or $\mathrm{Cyl}_{[3\bm{2}1]}$, respectively.)
It is unclear whether this branching prohibits orbits to
shadow sequences of the type~\eqref{heteroseq}.

The case $a < 1- 2/\sqrt{3}$ is simpler. If $|\Sigma_1|$ is sufficiently small,
no branching occurs
at the points $\{\mathrm{P}_{\langle 312 \rangle}\} \times \{\mathcal{Q}_{2/3}\}$
and $\{\mathrm{S}_{\langle 213 \rangle}\} \times \{\mathcal{T}_{3/2}\}$.
We conclude that there exist orbits in $\mathcal{X}$ that shadow
the sequence
\begin{equation}
\begin{CD}
\{\mathrm{S}_{\langle 213 \rangle}\} \times \{\mathcal{T}_{3/2}\} 
@<\text{detour}<<
\{\mathrm{S}_{\langle 213 \rangle}\} \times \{\mathcal{Q}_1\}    
@<\text{unique}<<   
\{\mathrm{P}_{\langle 312 \rangle}\} \times \{\mathcal{Q}_1\} 
@<\text{detour}<<
\{\mathrm{P}_{\langle 312 \rangle}\} \times \{\mathcal{T}_{3/2}\} 
\\
@V\text{detour}VV 
@.
@.
@AA\text{detour}A
\\
\{\mathrm{S}_{\langle 213 \rangle}\} \times \{\mathcal{Q}_{2/3}\} 
@>>\text{detour}>
\{\mathrm{S}_{\langle 213 \rangle}\} \times \{\mathcal{T}_1\} 
@>>\text{unique}> 
\{\mathrm{P}_{\langle 312 \rangle}\} \times \{\mathcal{T}_1\} 
@>>\text{detour}>
\{\mathrm{P}_{\langle 312 \rangle}\} \times \{\mathcal{Q}_{2/3}\} 
\end{CD}
\end{equation}
for some time in the asymptotic regime $\tau\rightarrow -\infty$.
Accordingly, if $|\Sigma_1|$ becomes sufficiently small in
the asymptotic evolution of a solution, then 
the solution enters an ``anti-LRS phase'' (where its behavior 
resembles the behavior of anti-LRS solutions).
The length of this phase (as measured, e.g., by the number
of oscillations in the sequence) is inversely proportional
to the initial value of $|\Sigma_1|$ and can thus be arbitrarily long.
During an anti-LRS phase, $|\Sigma_1|$ increases until either
the orbit branches off at $\{\mathrm{P}_{\langle 312 \rangle}\} \times \{\mathcal{Q}_{2/3}\}$
or $\{\mathrm{S}_{\langle 213 \rangle}\} \times \{\mathcal{T}_{3/2}\}$
when $|\Sigma_1|$ has become large enough, or
the orbit leaves the sectors $\langle 213 \rangle$ or $\langle 312 \rangle$.

It remains to ask whether more general considerations
than~\eqref{y2y3} can lead to a further exclusion of
parts of $\partial\mathcal{X}$ as possible
$\alpha$-limit sets.
For instance, it might turn out to be true that 
the $\alpha$-limit of a generic orbit in $\mathcal{X}$ is a subset of
the set 
\[ 
\bigcup_{ijk} \,\partial \mathrm{Cyl}_{[i\bm{j}k]}\:=\:
\big(\partial\mathit{\Sigma} \times \partial \mathcal{Y}\big)
\,\cup\,
\bigcup_{i} \Big( \overline{\mathit{\Sigma}} \times \{\mathcal{Q}_i\} \Big) 
\, \cup\,  
\bigcup_{i} \Big( \overline{\mathit{\Sigma}} \times \{\mathcal{T}_i\} \Big)\:.
\]

\noindent {\bf Acknowledgments:} S.C. is supported 
by FCT, Portugal  
(contract SFRH/BDP/21001/2004).


\begin{thebibliography}{99}

\bibitem{ABS} L.~Andersson, R.~Beig, B.~G.~Schmidt: 
Static self-gravitating elastic bodies in Einstein gravity.
{\it Commun.\ Pure Appl.\ Math.} (to appear).
Electronic archive: arXiv.org/gr-qc/0611108 (2007).

\bibitem{BS} R.~Beig, B.~G.~Schmidt: Relativistic elasticity. {\it Class. Quantum Grav.} {\bf 20}, 889--904 (2003)

\bibitem{CQ} B.~Carter, H.~Quintana: Foundations of general relativistic high-pressure elasticity theory. 
{\it Proc. R. Soc. Lond. A.} {\bf 331}, 57--83 (1972)

\bibitem{FHU} M.~Fj\"allborg, J.~M.~Heinzle, C.~Uggla: 
Self-gravitating stationary spherically symmetric systems in relativistic galactic dynamics.
{\it Math.\ Proc.\ Camb.\ Phil.\ Soc.} (to appear).
Electronic archive: arXiv.org/gr-qc/0609074 (2007).

\bibitem{HU} J.~M.~Heinzle, C.~Uggla: Dynamics of the spatially homogeneous Bianchi type I Einstein-Vlasov equations. 
{\it Class. Quantum Grav.} {\bf 23} No. 10, 3463--3489 (2006)

\bibitem{HUR} J.~M.~Heinzle, C.~Uggla, N.~R\"ohr: The cosmological billiard attractor. 
Electronic archive: arXiv.org/gr-qc/0702141 (2007) 

\bibitem{KM} J.~Kijowski, G.~Magli: Relativistic elastomechanics as a lagrangian field theory. 
{\it Journal Geom. Phys.} {\bf 9}, 207--223 (1992)

\bibitem{KS} M.~Karlovini, L.~Samuelsson: Elastic stars in general relativity: I. Foundations and equilibrium models. 
{\it Class. Quantum Grav.} {\bf 20} No. 16, 3613--3648 (2003)

\bibitem{MH} J.~E.~Marsden, T.~J.~R.~Hughes: {\it Mathematical Foundations of Elasticity}. 
Dover Publications, Inc. (1994)

\bibitem{P} J.~Park: Solutions of the Einstein Equations for Spherically Symmetric Elastic Bodies.
{\it Gen. Rel.\ Grav.} {\bf 32}, 235--252 (2000)
 
\bibitem{R} A.~D.~Rendall: The initial singularity in solutions of the Einstein-Vlasov system of Bianchi type I. 
{\it J. Math Phys.} {\bf 37}, 438--451 (1996)
 
\bibitem{T} A.~S.~Tahvildar-Zadeh: Relativistic and non-relativistic elastodynamics with small shear strains. 
{\it Ann. Inst. H. Poincar, Phys. thor.} {\bf 69}, No. 3, 275--307 (1998)

\bibitem{WE} J.~Wainwright, G.~F.~R.~Ellis: 
{\it Dynamical systems in cosmology}. Cambridge University Press, Cambridge (1997)

\bibitem{MWP} M.~M.~Wernig-Pichler: {\it Relativistic Elastodynamics}. PhD Thesis, University of Vienna.
Electronic archive: arXiv.org/gr-qc/0605025 (2006)

\end{thebibliography}
\end{document}